\documentclass{amsart}
\usepackage{amssymb}

\usepackage{graphicx,epstopdf}

\newcommand{\sgn}{{\rm{sgn\;}}}

\newcommand{\const}{{\rm{const\;}}}
\newcommand{\Cal}{\mathcal}

\newtheorem{theorem}{Theorem}
\newtheorem{lemma}{Lemma}
\newtheorem{prop}{Proposition}
\newtheorem{cor}{Corollary}

\newcommand{\hatQ}{{\hat Q}}
\newcommand{\dee}{{\mathrm d}}

\begin{document}

\title
[Spectra of $Sol$-manifolds]
{Spectra of $Sol$-manifolds: \\ arithmetic and quantum monodromy}
%
%
\author{A.V. Bolsinov}
\address{Department of Mathematics and Mechanics, Moscow State University, 119899 Moscow, Russia}
       \email{bolsinov@mech.math.msu.su}
\author{H.R. Dullin}
\address{Department of
Mathematical Sciences,
          Loughborough University,
Loughborough,
          Leicestershire, LE11 3TU, UK}
          \email{H.R.Dullin@lboro.ac.uk}
\author{A.P. Veselov}
\address{Department of
Mathematical Sciences,
          Loughborough University,
Loughborough,
          Leicestershire, LE11 3TU, UK
          and
Landau Institute for Theoretical Physics, Moscow, Russia}

\email{A.P.Veselov@lboro.ac.uk}

\begin{abstract}
The spectral problem of three-dimensional manifolds $M_A^3$
admitting $Sol$-geometry in Thurston's sense is investigated.
Topologically $M_A^3$ are torus bundles over a circle with
a unimodular hyperbolic gluing map $A$.
The eigenfunctions of the corresponding Laplace-Beltrami
operators are described in terms of the modified Mathieu functions. 
It is shown that the multiplicities
of the eigenvalues are the same for generic values of the parameters in the metric
and are directly related to the number of representations of an
integer by a given indefinite binary quadratic form. 
As a result the spectral statistics is shown to disagree with the Berry-Tabor conjecture.
The topological nature of the monodromy for both classical and 
quantum systems on $Sol$-manifolds is demonstrated.
\end{abstract}

\maketitle

\section
{Introduction}

It has been known since the nineteenth century that in dimension
two there is a close relationship between geometry and topology.
Namely each compact orientable surface admits a metric of constant
curvature: positive if it is a topological sphere, zero if it is a
torus and negative if it has genus more than 1.

In dimension three the situation is much more sophisticated. The
major development here was due to Thurston \cite{Th} who put
forward the famous {\it Geometrisation Conjecture:} any compact
orientable 3-manifold can be cut by disjoint embedded 2-spheres
and tori into pieces, which after gluing 3-balls to all boundary
spheres, admit one of 8 special geometric structures. These
special 3-dimensional geometries are the standard Euclidean $E^3$,
spherical $S^3$ and hyperbolic $H^3$ geometries, the product
geometries $S^2 \times {\mathbb R}$ and $H^2 \times {\mathbb R}$ and three
geometries related to the Lie groups $SL_2({\mathbb R})$, $Nil$ and
$Sol$.

The last group $Sol$ is the 3-dimensional solvable Lie group, which is isomorphic to the group of isometries of Minkowski 2-space.
The corresponding metric has the least symmetry
of all the 8 geometries as the identity component of the stabiliser of a point
is trivial.

The structure of 3-manifolds admitting any of the seven geometries
excluding the most complicated hyperbolic case $H^3$ is pretty
well understood. In particular a 3-manifold $M$ possesses
$Sol$-geometric structure if and only if $M$ is finitely covered
by a torus bundle over $S^1$ with hyperbolic gluing map. 
For all other 6 geometries $M$ must be a Seifert fibre space (see e.g.
\cite{Scott}), so the $Sol$-manifolds are special from this point of view.

Their special role in the theory of dynamical systems became clear after
a recent paper \cite{BT} by Taimanov and one of the authors, who 
showed the surprising fact that although the geodesic flow on
$Sol$-manifolds is integrable in the sense of Liouville 
(but not in the analytic category) it has non-zero topological entropy !

In the present paper we investigate the quantum version of
the geodesic flow on $Sol$-manifolds, which is the spectral
problem for the corresponding Laplace-Beltrami operator $\Delta.$
We describe the spectra explicitly in terms of the spectrum of the
modified Mathieu equation. These spectra are degenerate
and have very interesting arithmetic. The multiplicities are
directly related to the numbers of representations of a given
integer by an indefinite binary quadratic form determined by the
corresponding hyperbolic gluing map. This allows us to conclude that
the spectral statistics for  $Sol$-manifolds is not Poisson contrary to the 
well-known Berry-Tabor conjecture.


Note that the $Sol$-structure on $Sol$-manifolds is
not unique in the same way as the flat structure on a torus is. The
spectra of tori are very sensitive to a change of the flat
metric: if we change the periods slightly the degeneracy will
essentially disappear. The fact that this does not happen with
$Sol$-manifolds shows the {\it rigidity} of the spectra and can be considered as a reflection of the
hyperbolicity hidden inside the topology of $Sol$-manifolds.

We should mention that a deep relation of $Sol$-manifolds with arithmetic was known 
before (see e.g. \cite{ADS, Brezin, Hirz}).  In particular, Hirzebruch \cite{Hirz} and
 Atiyah, Donnelly and Singer \cite{ADS}  discovered a
remarkable relation between topological "signature defects" of $Sol$-manifolds
and arithmetical $L$-functions.

From the dynamical point of view the arithmetic and topology reveal themselves through {\it Hamiltonian monodromy} \cite{Duistermaat80}. Its quantum analogue - {\it quantum monodromy} - is a relatively new phenomenon \cite{CD, GU, San}, which still needs better understanding.  An interesting feature of our case is that the corresponding grid of the quantum states can be described explicitly and nicely visualised ({\it "$Sol$-flower"}, see fig.~\ref{solflower}, \ref{paralleltransport} below). This is probably the first example of quantum monodromy of that kind.


The structure of the paper is following. First we introduce the
class of $Sol$-manifolds and describe the classical geodesic
dynamics and the corresponding Hamiltonian monodromy. 
Then we review the 
facts from classical number theory about the relations between
binary quadratic forms and the modular group $SL(2, {\mathbb Z})$. In
section 5 we consider the spectral problem for the corresponding
Laplace-Beltrami operator and find the eigenfunctions in terms of
the modified Mathieu functions. The arithmetic of the multiplicities
of the eigenvalues is discussed in detail in section 6. The semiclassical analysis of the problem is done in section 7 in relation with Weyl's law. In section 8 we discuss the spectral statistics in the context of the Berry-Tabor conjecture \cite{Berry-Tabor}. The quantum monodromy for
$Sol$-manifolds is discussed in the final section. 

\section
{$Sol$-manifolds}

In this paper we restrict ourselves to the main class of $Sol$-manifolds,  which are $T^2$ torus 
bundles over a circle $S^1$ with hyperbolic gluing maps with positive eigenvalues. 
More precisely, consider the action of $\mathbb Z$ on $\tilde
M^3=T^2\times \mathbb R$ generated by the following transformation $T_A.$ Let $(x,y)$ be
standard periodic coordinates on $T^2$ defined modulo 1, and $z\in
(-\infty,+\infty)$ be a coordinate on $\mathbb R$. Then in these
coordinates the transformation $T_A$ is given by
\begin{equation}
T_A: \left(\begin{array}{cc} x \\ y \\ z \end{array}\right)
\longrightarrow \left(\begin{array}{cc} a_{11}x + a_{12}y \\
a_{21}x + a_{22}y
\\ z+1
\end{array}\right) \label{1}
\end{equation}
where $A=\left(\begin{array}{cc} a_{11} & a_{12} \\ a_{21} &
a_{22}
\end{array}\right) \in SL(2,{\mathbb Z})$ is an integer hyperbolic
matrix, which defines a hyperbolic automorphism of the 2-torus.
The corresponding $Sol$-manifold $M_A^3$ is defined as the quotient $\tilde M^3/\mathbb Z$
by this action.

Let $\lambda$ and $\lambda^{-1}$ be the eigenvalues of $A$ and we
assume that $\lambda > 1$. 
The $Sol$-manifolds with negative $\lambda$ are covered by those with positive eigenvalues.

Together with $(x,y,z)$ we shall use
another coordinate system $(u,v,z)$ on $M_A^3$, where $(u,v)$ are
linear coordinates on the fibres related to a positively oriented
eigenbasis of $A$. The transformation $T_A$ in these coordinates is given by
\begin{equation}
\left(\begin{array}{cc} u \\ v \\ z \end{array}\right) \longrightarrow
\left(\begin{array}{cc} \lambda u \\ \lambda^{-1} v \\ z+1 \end{array}\right)
\label{1'}
\end{equation}

One should note that unlike $(x,y)$, the new coordinates $(u,v)$
are not periodic on the tori $T^2$ anymore: two pairs $(u,v)$,
$(u',v')$ define the same point on $T^2$ if and only if
$(u-u',v-v')=k(c^1_1, c_1^2) + m(c^1_2, c_2^2)$, where  $k,m\in
\mathbb Z$ and $e_1=(c^1_1, c_1^2)$, $e_2= (c^1_2, c_2^2)$ is the
basis of the lattice $\Gamma$ associated to $T^2:$
$$
A=\left(\begin{array}{cc} a_{11} & a_{12} \\ a_{21} & a_{22}
\end{array}\right) =
\left(\begin{array}{cc} c^1_1 & c_2^1 \\ c^2_1 & c_2^2
\end{array}\right)^{-1} \left(\begin{array}{cc} \lambda & 0 \\ 0 &
\lambda^{-1} \end{array}\right) \left(\begin{array}{cc} c^1_1 &
c_2^1 \\ c^2_1 & c_2^2
\end{array}\right)
$$

The Riemannian metrics on $Sol$-manifolds come from right-invariant metrics 
on the universal
covering of $M_A^3,$ which has the natural structure of a solvable
Lie group $Sol.$ Topologically this group is ${\mathbb R}^3$ with a
multiplication of the form
$$
(u,v,w)*(u',v',w') = (u+e^{w}u',
v+e^{-w}v', w+w').
$$
One can realise it as the group of $3\times
3$ matrices of the form
$$
\left(\begin{array}{ccc} e^w & 0 & u
\\ 0 & e^{-w} & v \\
0& 0& 1
\end{array}\right).
$$

The $Sol$-manifolds $M_A^3$ we consider are the quotients of the
group $Sol$ by the discrete subgroups $G_A$ corresponding to $w=m
\ln \lambda, m \in {\mathbb Z}$ and $(u,v)= k e_1 + l e_2$ belonging
to the integer lattice $\Gamma$ described above, $z = w/ \ln \lambda.$

The right-invariant metrics on the group $Sol$ correspond to the following class of metrics on the 
$Sol$-manifold $M_A$:
\begin{equation}
\label{metric}
 \dee s^2=\alpha(z) \dee x^2+ 2\beta(z)\dee x\dee y +\gamma(z)\dee y^2 +
\dee z^2
\end{equation}
where
$$
\left(\begin{array}{cc} \alpha(z) & \beta(z)
\\ \beta(z) & \gamma(z)
\end{array}\right)=
 \exp(-zB)^\top
\left(\begin{array}{cc} \alpha & \beta \\ \beta & \gamma
\end{array}\right) \exp (-zB).
$$
Here $\alpha, \beta, \gamma$ are
real parameters with the only condition that
the form $\dee s^2=\alpha \dee x^2+ 2\beta \dee x \dee y +\gamma \dee y^2$ is positive
definite and $B$ is defined by the relation $\exp B=A:$
$$
B=
\left(\begin{array}{cc} c^1_1 & c_2^1 \\ c^2_1 & c_2^2
\end{array}\right)^{-1} \left(\begin{array}{cc}\ln \lambda & 0 \\ 0 &
- \ln \lambda \end{array}\right) \left(\begin{array}{cc} c^1_1 &
c_2^1 \\ c^2_1 & c_2^2
\end{array}\right).
$$
One can consider a more general metric allowing a constant coefficient at $\dee z^2$ but this will lead only to a general scaling.

\section{Geodesic flows on $Sol$-manifolds: integrals and Hamiltonian monodromy}

Thus, the Hamiltonian of the geodesic flow on $M_A^3$ in
$(u,v,z)$-coordinates can be written as
$$
H=\frac{1}{2} (E
e^{2z\ln\lambda}p_u^2+2Fp_up_v + Ge^{-2z\ln\lambda}p_v^2)+
\frac{1}{2} p_z^2,
$$
where $E,F,G$ are real parameters: $E >0$, $G >0$, $EG-F^2 >0$.
It is invariant under the following
transformation
\begin{equation}
T_A^*: \left(\begin{array}{cc}
u \\ v \\ z \\ p_u \\ p_v \\ p_z
\end{array}\right)
\longrightarrow
\left(\begin{array}{cc}
\lambda u \\ \lambda^{-1} v \\ z+1 \\ \lambda^{-1} p_u \\ \lambda p_v \\
p_z \end{array}\right)\,,
\label{2}
\end{equation}
and, of course, under the translations by the elements of the
lattice $\Gamma$. The same property must be satisfied for any smooth
function on $T^*M_A^3$, in particular, for the first integrals of the geodesic
flow.

Since $H$ depends neither on $u$, nor on $v$, the corresponding momenta
$p_u$ and $p_v$ are {\it local} first integrals of the geodesic flow.
However, being not invariant
under (\ref{2}), they are not well
defined on the cotangent bundle $T^*M_A^3$.
That is why, to get {\it global} first integrals, we need to replace $p_u$,
$p_v$ by two smooth functions $f_1(p_u,p_v), f_2(p_u,p_v)$ invariant under
 the transformation $(p_u,p_v)\to (\lambda^{-1} p_u , \lambda
p_v)$ (or, speaking in more general terms, by the invariants of the $\mathbb
Z$-action on the cotangent plane generated by the hyperbolic linear
transformation ${A^{\top}}^{-1}$).

One invariant function is evident:  $Q=p_up_v$. To find another
one we introduce the following expression which will be useful
also in the future
\begin{equation}
\label{alpha} \alpha = \frac{\ln\left( \sqrt{\frac{E}{G}} \left|
\frac{p_u}{p_v}\right|\right)}{2\ln\lambda}.
\end{equation}
Under the transformation (\ref{2}) $\alpha$ changes in a very
simple way:
$$
\alpha (p_u,p_v) \to \alpha(p_u, p_v) - 1
$$

Thus, as a second integral we can take any function of $\alpha$
with period 1, for instance, $\cos (2\pi\alpha)$ or $\sin
(2\pi\alpha)$.  However these functions are not smooth at $p_u = p_v = 0$. 
To avoid this difficulty and to get the first integrals in a more symmetric
form we put:
$$
\begin{array}{l}
 f_1=R(Q) \cos 2\pi\alpha,\\
 f_2=R(Q) \sin 2\pi\alpha,
\end{array}
$$
where
$$
R(Q)=\sqrt{|Q|} \exp (-\frac{1}{Q^2}).
$$

{\bf Remark.} The fact that the second integral is not analytic is
not accidental: the theorem proved by Taimanov \cite{T} implies
that $Sol$-manifolds do not admit integrable geodesic flows with
analytic integrals (see \cite{BT} for more details).

We are going to show now that one can see the topological
structure of the $Sol$-manifolds by looking at the Hamiltonian
monodromy of the geodesic flow. For that we will have to
investigate the bifurcation diagram (i.e.~the set of critical values) 
of the momentum mapping
restricted to the isoenergy surface $E^5_A=\{H=1\}$:
\begin{equation}
\Cal F_A=(f_1,f_2):E^5_A \to {\mathbb R}^2.
\label{3}
\end{equation}

\begin{prop}
The bifurcation diagram of the momentum mapping $\Cal F_A$ consists of two
circles
$$
f^2_1 + f_2^2 = R^2(Q^*_{\pm}),
$$
where
$Q^*_{\pm}=(F\pm \sqrt{EG})^{-1}$, and the point $(0,0)$, the centre of
these circles.  The set of critical points consists of five parts:
a) four
one-parameter families $L_i$ ($i=1,\dots,4$) of (degenerate) 2-dimensional
tori lying in the cotangent bundle and given by ($\alpha$ is a parameter):

\centerline{$z=-\alpha$, $u$ and $v$ are arbitrary,}

$$
p_z=0, \qquad
p_u=\pm\sqrt{\frac{e^{2\alpha\ln\lambda}}{E\left(1+\frac{F}{\sqrt{EG}}\right)}},
\qquad
p_v=\pm\sqrt{\frac{e^{-2\alpha\ln\lambda}}{G\left(1+\frac{F}{\sqrt{EG}}\right)}};
$$

and

\centerline{$z=-\alpha$, $u$ and $v$ are arbitrary,}

$$
p_z=0, \qquad
p_u=\pm\sqrt{\frac{e^{2\alpha\ln\lambda}}{E\left(1-\frac{F}{\sqrt{EG}}\right)}}
\qquad
p_v=\mp\sqrt{\frac{e^{-2\alpha\ln\lambda}}{G\left(1-\frac{F}{\sqrt{EG}}\right)}}
$$

b) the critical set $N$ given by the equation
$Q=p_up_v=0$.

\end{prop}

{\it Proof.}
We are interested in the singularities of $\Cal F_A$ or, which is the same,
those of the Liouville foliation. These singularities can be of two types.
To explain their nature we first consider the geodesic
flow on the covering manifold $\tilde M^3$. On this (non-compact) manifold
the integrals of the flow are simply $p_u$ and $p_v$.
Consider the Liouville foliation for this covering
system. Its singular leaves correspond to the critical points of the
momentum mapping
$$
\tilde{ \Cal F}=(p_u, p_v): \tilde E^5 \to {\mathbb R}^2,
$$
where $\tilde E^5=\{H=1\}\subset T^*\tilde M$.
Obviously, these leaves remain singular after the natural projection
$\tilde E^5 \to E^5_A$.  These are singularities of the first type.

On the other hand some new singularities appear since instead of $p_u$ and
$p_v$ we have to consider more complicated functions $f_1$ and $f_2$. In
other words, these are singularities of the map $(p_u, p_v) \to (f_1,
f_2)$.

Let us treat both cases in turns.
It is easily seen that $p_u$ and $p_v$ are functionally dependent, as
functions on $\tilde E^5=\{H=1\}$ if and
only if two conditions are simultaneously satisfied:  1)
$\frac{\partial H}{\partial p_z}=2p_z=0$ and 2)
$\frac{\partial H}{\partial z}=0$.  Taking into account the condition
$H=1$, we obtain a system of equations
$$
\begin{array}{l}
Ee^{2\log\lambda z}p_u^2- Ge^{-2\log\lambda z}p_v^2=0, \\ Ee^{2\log\lambda
z}p_u^2+2Fp_up_v+ Ge^{-2\log\lambda z}p_v^2=2 \end{array}
$$

The first equation gives
$$
z=-\frac{\ln\left( \sqrt{\frac{E}{G}} \left|
\frac{p_u}{p_v}\right|\right)}{2\ln\lambda}=-\alpha.
$$

Now solving this system with respect $p_u$ and $p_v$ (after substituting
$z=-\alpha$), we find four distinct solutions:
$$
\begin{array}{l}
 1)
\
p_u=\frac{e^{\alpha\ln\lambda}}{\sqrt{E\left(1+\frac{F}{\sqrt{EG}}\right)}},
\qquad
p_v=\frac{e^{-\alpha\ln\lambda}}{\sqrt{G\left(1+\frac{F}{\sqrt{EG}}\right)}};
\\
 2)
\
p_u=\frac{-e^{\alpha\ln\lambda}}{\sqrt{E\left(1+\frac{F}{\sqrt{EG}}\right)}},
\qquad
p_v=\frac{-e^{-\alpha\ln\lambda}}{\sqrt{G\left(1+\frac{F}{\sqrt{EG}}\right)}};
\\
 3)
\
p_u=\frac{e^{\alpha\ln\lambda}}{\sqrt{E\left(1-\frac{F}{\sqrt{EG}}\right)}},
\qquad
p_v=\frac{-e^{-\alpha\ln\lambda}}{\sqrt{G\left(1-\frac{F}{\sqrt{EG}}\right)}};
\\
 4)
\
p_u=\frac{-e^{\alpha\ln\lambda}}{\sqrt{E\left(1-\frac{F}{\sqrt{EG}}\right)}},
\qquad
p_v=\frac{e^{-\alpha\ln\lambda}}{\sqrt{G\left(1-\frac{F}{\sqrt{EG}}\right)}}.
\end{array}
$$

Thus, for each value of $\alpha$ we obtain four
2-dimensional invariant tori in $T^*M_A^3$. All of them are
diffeomorphically projected onto the same $T^2$-fibre
$T^2_{-\alpha}=\{z=\const=-\alpha\} \subset M$.
Varying $\alpha$, we obtain 4 families of
degenerate Liouville 2-tori $L_i$, $i=1,\dots, 4$.

It is easy to verify that for each family $L_i$ the value of $Q=p_up_v$
is constant and equal to   $Q^*_{+}=(F+\sqrt{EG})^{-1}$ for $L_1$ and
$L_2$, and equal to $Q^*_{-}=(F - \sqrt{EG})^{-1}$ for $L_3$ and $L_4$. Hence the
image of $L_1$ and $L_2$ is the circle  $f^2_1 + f_2^2 = R^2(Q^*_{+})$, and
analogously the image of $L_3$ and $L_4$ is the other circle
$f^2_1 + f_2^2 = R^2(Q^*_{-})$, as required.

The singularities of the second type come from those of the mapping
$(p_u, p_v)\to (f_1, f_2)$. It can be easily seen that the critical points
of this mapping are defined by the equation $Q=p_up_v=0$.
This implies immediately $f_1=f_2=0$ which gives
a single point on the bifurcation diagram, namely the centre of the
circles.

\begin{figure}
\centerline{\includegraphics[width=4cm]{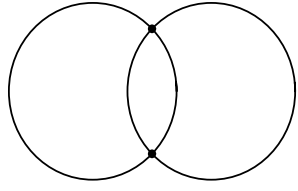}}
\caption{The topological structure of the singular leaf is $M^3_A \times K$ where $K$ is 
given in the figure.} \label{singularleaf}
\end{figure}

Notice that topologically the subset $N=\{ Q=p_up_v=0 \}\subset T^*M_A^3$ is
homeomorphic to the direct product $M_A^3 \times K$, where $K$ is a graph
that consists of two vertices and four segments connecting them (see
fig.~\ref{singularleaf}). This follows immediately from the parallelizability of
$M_A^3$ and the simple observation that in each cotangent space  the conditions $p_up_v=0$, $H=1$
define a graph homeomorphic to $K$.

Now we are able to describe the global structure of the foliation of the
isoenergy surface $E^5_A$ into Liouville tori.

If we remove the singular set from the isoenergy surface we obtain four
families of 3-dimensional Liouville tori distinguished from each
other by signs of $p_u$ and $p_v$:
$$
\begin{array}{l}
 a) \ \ p_u>0, p_v>0 ;\\
 b) \ \ p_u<0, p_v<0 ;\\
 c) \ \ p_u>0, p_v<0 ;\\
 d) \ \ p_u<0, p_v>0 .
\end{array}
$$

The families a) and b) are isomorphic (more precisely, they
transform into each other by the globally defined time reversal automorphism of
the geodesic flow $(u,v,z,p_u,p_v,p_z)\to
(u,v,z,-p_u,-p_v,-p_z)$).

The same is true for the families c) and d).

Each Liouville 3-torus is uniquely determined by the values of two
integrals $Q$ and $\alpha \bmod{1}$, where the values of $Q $ form the interval $(0,Q^*_+)$ in the first
two cases and $(Q^*_-, 0)$ for the other two cases. In particular, in each
of the cases, the base of the $T^3$-foliation is homeomorphic to a
punctured disc. As $Q\to 0$, the Liouville torus approaches the singular
set $N$. As $Q\to Q^*_\pm$, the torus shrinks into one of the degenerate
2-tori described above.

Thus, the base of the global Liouville foliation on $E^5_A=\{H=1\}$ can be
considered as four discs glued together at their centres. All interior
points of these discs except the centre correspond one-to-one to regular 3-dimensional
Liouville tori, the boundary circles of the discs correspond to the
families $L_i$ of degenerate 2-tori, and finally, the common center of the
discs corresponds to the singular set $N$.
\qed

The image of each family under the momentum map is a 2-disc with the center
removed.  This is exactly the situation when we can talk about Hamiltonian monodromy
\cite{Duistermaat80}.

\begin{theorem}
For each family of Liouville 3-tori there exist a basis of cycles in the first homology group of the tori  in which the Hamiltonian monodromy has the matrix
$$
\left(\begin{array}{cc} A & 0 \\ 0 & 1 \end{array}\right).
$$
\end{theorem}

{\it Proof.} This fact can be observed in many different ways. We shall
follows the definition of Hamiltonian monodromy and will explicitely
compute the deformation of Liouville tori and the final gluing map.

Consider an arbitrary Liouville 3-torus $T^3=T^3_{Q_0,\alpha_0}$.
In coordinates, this torus
is given by three conditions:
\begin{equation}
\begin{array}{l}
Ee^{2z\ln\lambda} p_u^2 +
2Fp_u p_v + Ge^{-2z\ln\lambda}p_v^2 + p_z^2=2, \\
Q(p_u, p_v)= p_up_v =
Q_0 \\
\alpha (p_u,p_v) = \frac{\ln\left( \sqrt{\frac{E}{G}} \left|
\frac{p_u}{p_v}\right|\right)}{2\ln\lambda} = \alpha_0 \mod 1
\end{array}
\label{81}
\end{equation}

More precisely, these conditions define a disjoint union of two or four
tori, which differ from each other by the signs of the momenta $p_u$ and $p_v$.
We consider one of them $T^3_{Q_0,\alpha_0}$ by putting for definiteness
$p_u>0$, $p_v>0$.

For our purposes first we need to explain why the above conditions
define indeed a three-dimensional torus and to describe the basic cycles
on this torus. Notice that the common level set (\ref{81}) of the first
integrals can be regarded from two slightly different points of view: as a
subset in $T^*\tilde M^3$ and that in $T^*M_A^3$. However one can show that the natural projection  $T^*\tilde M^3 \to
T^*M_A^3$ restricted to this level set is a diffeomorphism (no points are
glued between them). Thus, in fact there is no real difference between
these two points of view.  In particular, instead of conditions $p_up_v =
Q_0, \alpha (p_u,p_v) = \alpha_0 \mod 1$ we may simply assume that the
momenta $p_u$, $p_v$ themselves are constant.  Then the conditions
(\ref{81}) can be rewritten as:
$$
p_u=\const, \quad p_v=\const, \quad
\hbox{$u$ and $v$ are arbitrary,}
$$
and
\begin{equation}
c_1 \cosh
(2\ln\lambda (z+\alpha_0)) + p_z^2 = c_2,
\label{91}
\end{equation}
where
$c_1=2\sqrt{EG}|Q_0|$, $c_2=2-2FQ_0$. We see that the variables separate
and the fact that this systems defines a 3-torus becomes evident.  Indeed,
the variables $u,v$ "run" over a two-dimensional torus and the last
equation defines a simple closed curve on the plane ${\mathbb R}^2(z,p_z)$.
In other words, we have a natural splitting of $T^3_{Q_0,\alpha_0}$ into
the direct product $T^2 \times S^1$.  Thus, as basic cycles on
$T^3_{Q_0,\alpha_0}$ we can take the cycles on
$T^2(u,v)$ related to the original coordinate system $(x,y)$ (see above)
and the third cycle defined by (\ref{91}).

Now let us look at what happens to this torus if we change the parameters
$Q_0$ and $\alpha_0$ in such a way that the point $\Cal
F_A(T^3_{Q_0,\alpha_0})$ moves inside the image of the momentum mapping
around the singular point $\Cal F_A(N)=(0,0)$.  It is easy to see that
this deformation just means that we change the value of $\alpha$, while
$Q$ can be chosen to remain constant:
$$
Q(t)= Q_0, \qquad \alpha(t)=\alpha_0+ t, \qquad
t\in [0,1].
$$

Consider the family of mappings
$$
\phi_t(u,v,z,p_u,p_v,p_z) = (u,v, z-t, e^{t\ln\lambda} p_u,
e^{-t\ln\lambda} p_v, p_z).
$$

It is not hard to see that the image of $T^3_{Q_0,\alpha_0}$ under $\phi_t$
is exactly $T^3_{Q_0,\alpha_0+t}$ and
$\phi_t: T^3_{Q_0,\alpha_0} \to T^3_{Q_0, \alpha_0+t}$ is a difeomorphism.
In other words, $\phi_t$ defines the deformation of Liouville tori we need.

At the moment $t=1$ the torus comes back to the initial position, i.e.,
$T^3_{Q_0,\alpha_0} = T^3_{Q_0, \alpha_0+1}$, and we obtain the monodromy
map
$$
\phi_1: T^3_{Q_0,\alpha_0} \to T^3_{Q_0,\alpha_0}=T^3_{Q_0, \alpha_0+1}
$$

Now our goal is to describe the corresponding automorphism of the first
homology group:
$$
{\phi_1}_* : H_1(T^3_{Q_0,\alpha_0} )={\mathbb Z}^3 \to
H_1(T^3_{Q_0,\alpha_0})={\mathbb Z}^3.
$$

Using the identification (\ref{2}) we see that the map $\phi_t$ can be
rewritten as follows:
$$
\phi_t \left(\begin{array}{c} u \\ v \\ z \\ p_u
\\ p_v \\ p_z \end{array}\right)                     =
\left(\begin{array}{c} \lambda u \\ \lambda^{-1} v \\ z \\ p_u \\ p_v \\
p_z \end{array}\right)
$$

We see that the only transformation is related to the variables $u$ and
$v$. Moreover, this transformation is exactly the original hyperbolic
automorphism $A: T^2 \to T^2$. Taking into account the natural splitting
$T^3_{Q_0,\alpha_0}=T^2(u,v)\times S^1(z,p_z)$ we conclude immediately
that the monodromy matrix in the chosen basis is
$$
\left(\begin{array}{cc} A & 0 \\ 0 & 1 \end{array}\right) \,.
$$
\qed

We conclude this section with the discussion of the geodesics on
$Sol$-manifolds. They have different properties depending on the
types of leaves of the Liouville foliation which they belong to.

First consider the geodesics lying on Liouville tori of
dimension three. They are characterized by the property that all
momenta $p_u, p_v$ and $p_z$ differ from zero. More precisely,
the signs $p_u$ and $p_v$ always remain the same, whereas the sign
of $p_z$ changes. This happens when $z$ reaches the value
$$
z_\pm=\frac {\pm \cosh^{-1} {\left( \frac {h-F p_u p_v}
{\sqrt{EG}|p_up_v|} \right) } } {2\ln\lambda} - \alpha(p_u,p_v).
$$

Two levels $z=z_+$ and $z=z_-$ are exactly the caustics of the
Liouville tori that contains a given geodesic. The situation is
quite similar to that on a surface of revolution where
the motion takes place between two levels of $z$.

It is easy to see that the distance between these levels $z_+ -
z_-$ tends to infinity as $p_up_v$ tends to zero. From this it
follows that the corresponding geodesics rotate many times (along
the base $S^1$), then turn back, after this go in the opposite
direction, then turn back and so on. As $p_up_v$ tends to zero the
number of rotations in one direction until turning back (or, which
is the same, between two caustics) increases up to infinity.

If $p_up_v=0$, then we are on the singular level. The
corresponding geodesics have the following behaviour. If both
$p_u$ and $p_v$ vanish, then we obtain the family of geodesics
$$
u=const, \ \
v=const, \ \ z= t .
$$

Such geodesics obviously form an invariant submanifold $N_+$ in
$T^*M$ which is diffeomorphic to $M$. Exactly on this submanifold
the geodesic flow is chaotic and has positive entropy. Indeed, the
time-one map transform each fibre $T^2_z$ into itself by means of
the hyperbolic automorphism $A$. As well known, the entropy of
$A:T^2\to T^2$ is $\ln\lambda>0$.

There is another invariant submanifold $N_-$ with the same
properties formed by vertical geodesics going in the opposite
direction:
$$ u=const, \ \ v=const,  \ \ z= -t .
$$

From the viewpoint of the ambient geodesic flow $N_+$ and $N_-$
are hyperbolic invariant subsets. The stable manifold coresponding
to $N_+$ is given by $p_v=0$, the unstable one is $p_u=0$. For
$N_-$ the stable and unstable manifolds interchange. The geodesics
satisfying the condition $p_v=0$ as $t\to +\infty$ asymptocally
approaches $N_+$, in particular, $p_z \to +1$. But there is
$t=t_0$ when $p_z$ changes sign so that for $t\to -\infty$ the
geodesic approaches to $N_-$. The geodesics satisfying $p_u=0$
behave in the opposite way.

In slightly other terms this structure can be described as
follows: there are two hyperbolic submanifolds diffeomorphic to
$M_A^3$, they are  connected by 4 four-dimensional separatrices,
see fig.~\ref{singularleaf}

Finally we would like to mention an interesting phenomenon
which one would not expect from an integrable geodesic flow on a compact manifold.
Namely, one of the action integrals diverges as the integral $Q \to 0$ with the energy fixed
(see the calculations and footnote in Section 7). Normal scenario would be when approaching the singular level some of the cycles of the Liouville tori shrink so the actions will stay finite.
The fact that this not true for $Sol$-manifolds when one approaches the singular (chaotic) level 
demonstrates once again the peculiar nature of this system.

To discuss the quantum case we will need some facts from
the classical number theory, which we present in the next section.

\section{$SL(2,{\mathbb Z})$ and binary quadratic forms}

The content of this section is well-known (see e.g.
\cite{Landau,LeVeque,Sarnak}).

Let $A=\left(\begin{array}{cc} a_{11} & a_{12} \\ a_{21} & a_{22}
\end{array}\right) \in SL(2,{\mathbb Z})$ be an integer hyperbolic matrix. Hyperbolicity
as before means that its eigenvalues are real and distinct. We
would like to consider $A$ as the automorphism of the lattice
${\mathcal L} = {\mathbb Z} \oplus {\mathbb Z} \in {\mathbb R}^2$ by choosing
some basis $e_1, e_2$ in this lattice.

For any such $A$ we can define the following integer binary
quadratic form $Q_A$ by the formula
\begin{equation}
\label{QA} A {\bf v}\wedge {\bf v} = Q_A ({\bf v}) e_1 \wedge e_2,
\end{equation}
where ${\bf v}$ is a vector from ${\mathbb R}^2.$ Explicitly if ${\bf
v} = x e_1 + y e_2$ then
\begin{equation}
\label{QAexp} Q_A(x,y) = \det\left(\begin{array}{cc} a_{11}x
+a_{12}y & x
\\ a_{21}x +a_{22}y & y \end{array}\right) = -a_{21} x^2 + (a_{11}-a_{22}) xy + a_{12} y^2.
\end{equation}
It is easy to see from the definition that this form is invariant
under the action of $A$:
$$
Q_A(A{\bf v}) = Q_A({\bf v}).
$$

Notice that $Q_A$ has the discriminant
$$
D = (a_{11}-a_{22})^2 +
4a_{12}a_{21} = (a_{11}+a_{22})^2 - 4(a_{11}a_{22}-a_{12}a_{21})=
(a_{11}+a_{22})^2 -4,
$$
which is exactly the discriminant of the
characteristic equation of $A$:
$$
\lambda^2 - (a_{11}+a_{22})
\lambda + 1 = 0.
$$
In particular, since $A$ is hyperbolic the form
$Q_A$ is indefinite. Note that the discriminant $D$ cannot be a total square.

In general the coefficients of the quadratic form $Q_A$ may have a
common factor. Let
\begin{equation}
\hatQ_A(x,y)= a x^2 + b xy + c y^2
\end{equation}
be its {\it primitive form} after division of $Q_A$ by the largest
common factor. It is defined correctly only up to a sign.

Thus to each integer unimodular hyperbolic matrix $A$ we relate an
indefinite integer primitive quadratic form $\hatQ_A$.

Conversely, suppose we have such a form $Q(x,y)= a x^2 + b xy + c
y^2.$ We would like to describe all $A$ from $SL(2,{\mathbb Z})$
which preserve this form. Such $A$ are called the {\it automorphs}
of $Q.$ Let
$$
d = b^2 - 4ac
$$
be the discriminant of $Q$ which we
assume not to be a total square and consider the corresponding
Diophantine equation called {\it Pell's equation}:
\begin{equation}
\label{Pell} X^2 - dY^2 =4.
\end{equation}
Then the group of automorphs consists of matrices of the
form
$$
A=\pm \left(\begin{array}{cc} \frac{X-bY}{2} & -cY \\ a Y &
\frac{X+bY}{2}
\end{array}\right),
$$
where $(X,Y)$ are the solutions of Pell's equation. Modulo
$\pm I$ this group is cyclic with generator
\begin{equation}
\label{A0} A_0 = \left(\begin{array}{cc} \frac{X_0-bY_0}{2} &
-cY_0
\\ a Y_0 & \frac{X_0+bY_0}{2}
\end{array}\right),
\end{equation}
where $(X_0,Y_0)$ is the fundamental solution of this equation.

Recall that $(X_0,Y_0)$ is the {\it fundamental solution} of
Pell's equation if $X_0 >0, Y_0 >0$ and $X_0 + \sqrt d Y_0$ is
minimal among all such solutions. 
The classical result about
Pell's equation says that all other solutions can be found from
the relation
$$
\frac{X+\sqrt{d}Y}{2} = \pm
\left(\frac{X_0+\sqrt{d}Y_0}{2}\right)^n,
$$
where $n=0,1,\dots.$ One can find the fundamental solution from the continued fraction of $\sqrt{d}$.
This structure of the solutions of Pell's equations induces the cyclic group structure
for the automorphs.

Notice that the form $Q_A$ corresponding to the matrix (\ref{A0})
has the form
$$
Q = Y_0 (a x^2 + b xy + c y^2).
$$

Let us call a hyperbolic element $A$ from $SL(2,{\mathbb Z})$ {\it
primitive} if it can not be represented as a power of any other
element from $SL(2,{\mathbb Z})$.

Thus we have described a natural correspondence between the
primitive binary indefinite forms $Q$ and primitive elements $A$
from $SL(2,{\mathbb Z}).$ In particular, it helps us to answer the
question if a given integer unimodular matrix $A$ is a primitive
or if not which power of a primitive matrix it is.

\section{Spectrum and eigenfunctions of the Laplace-Beltrami operator}

Let us now discuss the quantum geodesic problem on the
$Sol$-manifold $M_A^3$:
\begin{equation}
\label{eigen} -\Delta \psi  = \mathcal E \psi,
\end{equation}
where $\Delta$ is the Laplace-Beltrami operator on $M_A^3$ and $\psi
= \psi(P, \mathcal E), P \in M_A^3.$ In coordinates $(u,v,z)$ the Laplace-Beltrami
operator has the following explicit form:
\begin{equation}
\Delta = E e^{2z\ln\lambda}\frac{\partial^2}{\partial u^2}+
2F\frac{\partial^2}{\partial u\partial v} +
Ge^{-2z\ln\lambda}\frac{\partial^2}{\partial v^2}+
\frac{\partial^2}{\partial z^2}.
\label{3'}
\end{equation}
This is a self-adjoint operator in the Hilbert space $L_2(M_A^3)$ where
the integration measure on $M_A^3$ is induced by the Riemannian
metric (\ref{metric}). In both $(x,y,z)$ and $(u,v,z)$ coordinate
systems the corresponding measure $d\mu$ is proportional to the
standard Lebesgue measure on ${\mathbb R}^3.$

Because the coefficients of $\Delta$ depends only on $z$ it is quite natural to separate variables and look for the eigenfunctions of $\Delta$ of the form
$$
\Psi_\gamma(u,v,z)=e^{2\pi i (\gamma , w)} f(z),
$$
where
$\gamma$ is an element of the dual lattice $\Gamma^*$
corresponding to the $T^2$-fibres and $w=(u,v)$ (so the scalar product $(\gamma , w)$ is
defined modulo $\mathbb Z$).

By substituting into the Schr\"odinger equation (\ref{eigen}), (\ref{3'}) we get
$$
\Delta \Psi_\gamma = \left(
\frac{\partial^2 f}{\partial z^2} - 8\pi^2 \sqrt{EG} |Q(\gamma)|
\left( \cosh \bigl(2\ln\lambda (z+\alpha(\gamma) ) \bigr) +
\frac{F  \sgn Q(\gamma)}{\sqrt{EG}} \right) f \right) e^{2\pi i
(\gamma , w)},
$$
where $Q(\gamma)=(\gamma, e_u)(\gamma, e_v)$ is
a quadratic form on the lattice $\Gamma^*$, and
$$
\alpha(\gamma)=\frac{\ln{\left( \sqrt{\frac{E}{G}} \left|
\frac{(\gamma,e_u)}{(\gamma, e_v)}\right|
\right)}}{2\ln\lambda}.
$$

Here $e_u$ and $e_v$ are the eigenvectors of $A$ related to the
eigenvalues $\lambda$ and $\lambda^{-1}$ respectively and the
basis $e_u, e_v$ is assumed to be positively oriented. Notice that
$\alpha$ is the same as before in (\ref{alpha}) if we replace
$p_u$ by $(\gamma, e_u)$ and $p_v$ by $(\gamma, e_v).$

To clarify the meaning of the coefficient in front of the $\cosh$
let us consider the basis $e_u^*, e_v^*$ in ${\mathbb {R}}^{2*}$ dual to
$e_u, e_v$. The vectors $e_u^*$ and $e_v^*$ are also the
eigenvectors of $A^*$ with the eigenvalues $\lambda$ and
$\lambda^{-1}$ respectively. By definition we have
$\gamma=(\gamma,e_u)e_u^* + (\gamma,e_v)e_v^*$. Since $Q(\gamma)$
is obviously invariant under the action of $A^*$ it is natural to
compare it with the binary form $Q_{A^*}$ defined in the section
3. We have
\begin{align*}
A^*\gamma \wedge \gamma & = (\lambda (\gamma,e_u)e_u^*
+ \lambda^{-1}(\gamma,e_v)e_v^*) \wedge((\gamma,e_u)e_u^* +
(\gamma,e_v)e_v^* )  \\
& =
(\gamma,e_u)(\gamma,e_v)(\lambda-\lambda^{-1}) e_u^*\wedge e_v^*.
\end{align*}
Let $l_1,l_2$ be a positively oriented basis in the dual
lattice $\Gamma^*,$ then by definition $A^*\gamma \wedge \gamma =
Q_{A^*}(\gamma) l_1 \wedge l_2.$

From these calculations and from the equalities $E= |e_u^*|^2, G =
|e_v^*|^2$ it follows that
$$
\sqrt{EG} |Q(\gamma)| = c
|Q_{A^*}(\gamma)|,
$$
where
\begin{equation}
\label{c} c = c(A; E,F,G) = \frac{{\mathcal A}(\Box^*)}{\sqrt D
\sin \theta} = \frac{1} {\sqrt D {\mathcal A}(T^2)
\sin \theta}
\end{equation}
${\mathcal A}(\Box^*)$ is the area of the dual basic parallelogram
$\Pi(e_1^*,e_2^*)$ (which is the inverse of the area of the fibre $T^2$), $D=(\lambda-\lambda^{-1})^2$ is the discriminant
of the characteristic equation of the matrix $A$ (or equivalently
$A^*$), and $\theta$ is the angle between $e_u^*$ and $e_v^*$.
Thus we have proved the following
\begin{prop}
A function $\Psi = e^{2\pi i (\gamma , w)} f(z)$ satisfies
equation (\ref{eigen}) if and only if $f(z)$ satisfies the
modified Mathieu equation
\begin{equation}
\label{cM} \left( -\frac{\dee^2}{\dee z^2}  + |\nu(\gamma)| \cosh 2\mu
(z+\alpha(\gamma)\right ) f(z) = \Lambda f(z),
\end{equation}
where $\mu = \ln \lambda, \,\, \nu(\gamma) = 8 \pi^2 c
Q_{A^*}(\gamma)$ and $\alpha(\gamma)$ is given above. The
eigenvalues $\mathcal E$ and $E$ are related by the shift
\begin{equation}
\label{shift} \mathcal E = \Lambda + \nu(\gamma) \cos \theta.
\end{equation}
\end{prop}

Recall that the {\it modified Mathieu equation} is the $\cosh$-version of the standard Mathieu equation
$$\frac{\dee^2 y}{\dee x^2}  +  (a \cos 2 \mu x + b) y  = 0.$$ Its solutions are known as {\it modified Mathieu functions} (see e.g. \cite{WW,Z}). They appear also 
in the theory of {\it Coulomb spheroidal functions} \cite{KPS}, where one can find some related numerical results (see also \cite{KTsi}).

Let $ \Lambda =  \Lambda_k (\nu)$, $k = 1,2,\dots$ be the spectrum of the
corresponding modified Mathieu operator 
$$
\mathcal M = -\frac{\dee^2}{\dee z^2}  + |\nu| \cosh 2\mu z 
$$
and $f_{\gamma,k}(z)$ be the corresponding solutions of (\ref{cM}). 

Thus, to each element $\gamma$ of the dual lattice $\Gamma^*$ we associate the functions
$\Psi_{\gamma,k}(u,v,z)=e^{2\pi i (\gamma , w)}
f_{\gamma,k}(z).$
The problem with these functions is that they are well defined on
the covering space $\tilde M^3 = T^2\times \mathbb R$ but not on the
$Sol$-manifold $M_A^3$ itself because they are not invariant with
respect to the transformation (\ref{1}), (\ref{1'}). One can try
to construct the genuine eigenfunctions of $\Delta$ on $M_A^3$ by
averaging these functions with respect to the action of $\mathbb Z$
on $\tilde M^3$ generated by this transformation. It turns out
that the averaging procedure works.

To show this let us consider instead of $\Psi_{\gamma,k}(u,v,z)$
the following sum
\begin{equation}
\label{Phi} \Phi_{\gamma,k} = \sum _{n\in\mathbb Z}
\Psi_{\gamma,k}(\lambda^n u, \lambda^{-n} v, z+n) = \sum
_{n\in\mathbb Z} \Psi_{{A^*}^n\gamma,k}(u, v, z).
\end{equation}
Because of the fast decay of the eigenfunctions $f_{\gamma,k} (z)$
this sum is absolutely convergent. It is easy to see that it
defines a well-defined function on $M_A^3,$ which is an
eigenfunction of the Laplace-Beltrami operator $\Delta.$

The eigenfunctions $\Phi_{\gamma,k}(u,v,z)$ on $M_A^3$ actually
depend only on the orbits $[\gamma]=\{ {A^*}^n(\gamma)\}_{n\in\mathbb
Z}$ with respect to the action of $A^*$ on $\Gamma^*$:
$\Phi_{\gamma,k}(u,v,z) = \Phi_{[\gamma],k}(u,v,z)$.

We should also consider separately the eigenfunctions related to
$\gamma=0$. It is easy to see that the corresponding
eigenfunctions have the very simple form
\begin{equation}
\Phi_{0,s} = 1, \cos {2\pi z}, \sin {2\pi z}, \cos {4\pi z},
\sin{4\pi z}, \dots, \cos{2k\pi z}, \sin{2k\pi z}, \dots \label{6}
\end{equation}
with the eigenvalues $\mathcal E_k = (2 \pi)^2 k^2.$

\begin{theorem}
The eigenfunctions of the Laplace-Beltrami operator
$\Phi_{[\gamma],k}(u,v,z), [\gamma] \in \Gamma^*\setminus \{0\} / {A^*} $ and
$\Phi_{0,s}(z)$ form a complete basis in
$L_2(M_A^3)$.
\end{theorem}

{\em Proof}. The independence and orthogonality of these functions are
obvious. The only thing we have to verify is the completeness. To
prove this we need to show that any smooth function $\Phi: M_A^3\to
\mathbb R$ which is orthogonal to each eigenfunction from the list
is, in fact, zero.

Consider such a function $\Phi(w,z)$ on the covering space $\tilde
M^3$ and expand it as a Fourier series (with respect to $w$):
$$
\Phi(w,z)=\sum_{\gamma\in\Gamma^*} e^{2\pi i (\gamma,w)}
a_\gamma(z)
$$
with some smooth coefficients $a_\gamma(z), z \in {\mathbb R}.$

\begin{lemma}
For all $\gamma\ne 0$ the functions $a_\gamma(z)$ have fast
decay at infinity and thus belong to $L^2(\mathbb R).$
\end{lemma}

{\em Proof}. Since $\Phi$ is invariant with respect to the transformation
(\ref{1}), we have $\Phi(w,z)=\Phi(Aw,z+1)$. Hence
$$
\sum_{\gamma\in\Gamma^*} e^{2\pi i (\gamma,w)} a_\gamma(z)=
\sum_{\gamma\in\Gamma^*} e^{2\pi i (\gamma,Aw)} a_\gamma(z+1)=
\sum_{\gamma\in\Gamma^*} e^{2\pi i (A^*\gamma,w)} a_\gamma(z+1).
$$
Thus the Fourier coefficients satisfy the following property:
$$
a_{\gamma}(z+1) = a_{A^*\gamma}(z),
$$
or, more generally,
$$
a_{\gamma}(z+n)=a_{{A^*}^n\gamma}(z), \qquad n\in\mathbb Z.
$$
Since
the Fourier coefficients $a_\gamma$ of a smooth function decay
fast for large $\gamma$ and ${A^*}^k\gamma$ for $\gamma\ne 0$ tends
to infinity we see that the functions $a_{\gamma}(z)$ decay very
fast and thus belong to $L^2(\mathbb R)$. 
\qed

Now suppose that $\Phi(w,z)$ is orthogonal to the eigenfunction
$\Phi_{[\gamma_0],k}(u,v,z)= \sum _{n\in\mathbb Z}
\Phi_{{A^*}^n\gamma_0,k}(u, v, z).$ Since the measure on $M_A^3$ is
proportional to the standard Lebesgue measure $\dee u \dee v \dee z$ we have
\begin{align*}
0&=\langle \Phi(w,z), \Phi_{[\gamma_0],k}(w,z)\rangle = \int_{M_A^3}
\Phi(w,z)\bar\Phi_{[\gamma_0],k}(w,z)\dee\sigma\\
&= 
\int_0^1 \left(
\sum_{\gamma\in\Gamma^*}\sum _{n\in\mathbb Z}\int_{T^2}
 e^{2\pi i (\gamma,w)}
e^{-2\pi i ({A^*}^n\gamma_0 , w)}
\dee u \dee v \right) a_\gamma(z) f_{{A^*}^n\gamma_0,k}(z) \dee z \\
&=
\int_0^1 \left( \sum _{n\in\mathbb Z}\int_{T^2}
 e^{2\pi i ({A^*}^n\gamma_0,w)}
 e^{-2\pi i ({A^*}^n\gamma_0 , w)}
\dee u\dee v \right) a_{{A^*}^n\gamma_0}(z) f_{{A^*}^n\gamma_0,k}(z) \dee z \\
&=
{\mathcal A}{(T^2)}\int_0^1  \sum _{n\in\mathbb Z}
 a_{{A^*}^n\gamma_0}(z) f_{{A^*}^n\gamma_0,k}(z) \dee z \,.
\end{align*}
We now use the property that $
f_{\gamma,k}(z+n)=f_{{A^*}^n\gamma,k}(z)$ and $
a_{\gamma}(z+n)=a_{{A^*}^n\gamma}(z),\quad n\in\mathbb Z$ to conclude
that
\begin{align*}
\int_0^1  \sum _{n\in\mathbb Z}
 a_{{A^*}^n\gamma_0}(z) f_{{A^*}^n\gamma_0,k}(z) \dee z
& = \\
\int_0^1  \sum _{n\in\mathbb Z}
 a_{\gamma_0}(z+n) f_{\gamma_0,k}(z+n) \dee z & =
\int_{-\infty}^{+\infty}a_{\gamma_0}(z) f_{\gamma_0,k}(z) \dee z.
\end{align*}
Thus, the Fourier coefficients $a_{\gamma_0}(z)$  for
$\gamma_0\ne 0$ belong to $L^2(\mathbb R)$ and at the same time are
orthogonal to all the functions $f_{\gamma_0,k}(z)$ which form a
complete basis in $L^2(\mathbb R)$. Hence for $\gamma_0\ne 0$ the
coefficients $a_{\gamma_0}(z)\equiv 0$.

This means that the function $\Phi$ must be of the form
$\Phi(w,z)=a(z),$ where $a(z)$ is periodic with period 1. Now
using orthogonality to the functions (\ref{6}) we conclude that
$a(z)$ must be identically zero. 
\qed

\begin{cor}
The spectrum of the Laplace-Beltrami operator on
$Sol$-manifolds consists of two parts: the trivial part
$$
\mathcal E =
\mathcal E_k = 4 k^2 \pi^2,\quad k=0,1,\dots
$$
corresponding to
the eigenfunctions (\ref{6}) and the non-trivial part
$$
\mathcal E =
\mathcal E_{l, [\gamma]} =  \Lambda_l (\nu([\gamma])) + \nu([\gamma])
\cos \theta, \quad l = 1,2 \dots, \quad [\gamma] \in \Gamma^*\ \{0\}
/{A^*}
$$
related to the modified Mathieu equation
(\ref{cM}).
\end{cor}

The multiplicities of the trivial eigenvalues are 2 except for the
ground state $\mathcal E = 0$ which has multiplicity 1. The
multiplicities of the non-trivial part of the spectrum are much
more interesting and the answer depends on the arithmetical
properties of the gluing map $A.$ We discuss this in the next
section.

\section{Multiplicities of the eigenvalues and number theory}

\begin{figure}
\centerline{ \includegraphics[width=8cm]{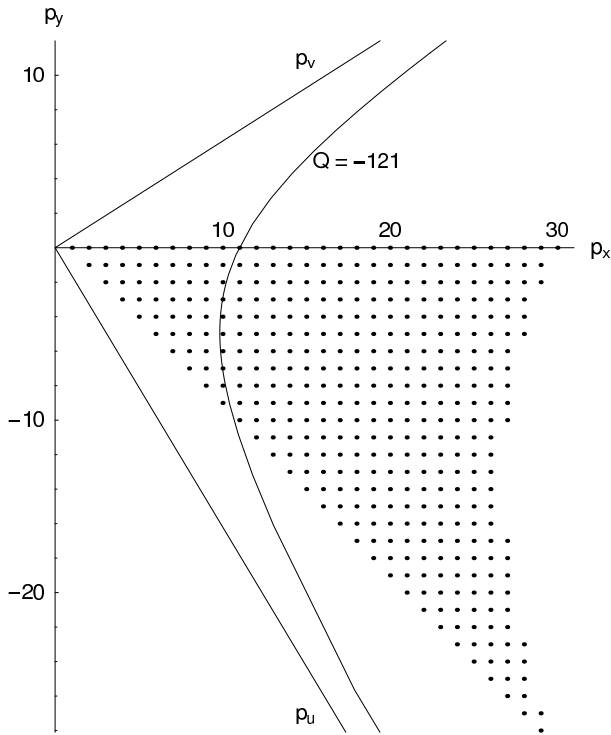} }
\caption{Fundamental domain of the lattice in $(p_x, p_y)$ with $|Q| \le 30^2$ for the cat-map.
The hyperbola $Q = -11^2$ illustrates the first example of a non-trivial degeneracy.} \label{lattice}
\end{figure}

As one can see from the previous section the eigenvalue of
$\Phi_{[\gamma],k}(u,v,z)$ depends on $\gamma$ only via $Q_{A^*}(\gamma)$. Thus
the calculation of the multiplicity (for generic values of the parameters to avoid additional accidental coincidences) is reduced to the classical
number theoretic problem of finding the number $N_Q(n)$ of integer
solutions of the equation $Q(x,y) = ax^2 + bxy + cy^2 = n$ for a
primitive indefinite quadratic form $Q$ different modulo its
automorphs. Figure 3 illustrates this for the cat-map $A$ with $Q = -x^2 +xy +y^2.$

For forms $Q$ with certain discriminants there exists an effective formula
which allows to compute $N_Q(n)$ 
To be more precise we need the following notion. We say that two
forms $Q$ and $Q'$ are equivalent if there exists a transformation
from $SL(2,{\mathbb Z})$ mapping one into another. It is easy to see
that two equivalent forms must have the same discriminant $d =
b^2-4ac.$ The converse is not true: there can be more than one
non-equivalent forms with the same discriminant.

Let $h(d)$ be the number of classes of primitive forms with
discriminant $d.$ Note that the discriminant $d = b^2 - 4 ac $ is always
0 or 1 modulo 4 and we assume as usual that it is not a total
square. 

{\bf Remark.} One should distinguish $h(d)$ and the class number of
ideals in the quadratic number field ${\bf Q}(\sqrt{d}).$ They
coincide only if the so-called {\it negative Pell equation}
$$
X^2 - dY^2 = -4
$$
has a solution; otherwise $h(d)$ is twice
as big (see e.g. \cite{Hua}, Chapter 16). The last property can be
reformulated in terms of the period of the continued fraction
expansion of $\sqrt{d},$ but a more explicit description is unknown.

If $h(d) = 1$ then all forms with the
discriminant $d$ are equivalent. In that case there is the
following remarkable formula for the number $N_Q(n)$ when $n$ is positive and 
coprime with $d$:
\begin{equation}
\label{Nn} N_Q(n) = N_d(n) = \sum_{k|n} \left(\frac{d}{k}\right),
\end{equation}
where the sum is taken over all divisors of $n$ and
$\left(\frac{d}{k}\right)$ is the standard Kronecker symbol (see Landau 
\cite{Landau}, Chapter IV.4). The {\it Kronecker symbol} is a
real character modulo $d$, which has the following properties
determining it uniquely:
\begin{enumerate}

\item If $d$ and $k$ are not coprime then $\left(\frac{d}{k}\right)=0;$

\item If $d$ and $k$ are coprime then $\left(\frac{d}{k}\right)= \pm 1;$

\item $\left(\frac{d}{kl}\right)=\left(\frac{d}{k}\right)\left(\frac{d}{l}\right)$

\item for $p$ odd prime which is not a divisor of $d$
$\left(\frac{d}{p}\right)$ coincides with the {\it Legendre
symbol}, which is $1$ if $d$ is quadratic residue modulo $p$ and
$-1$ otherwise;

\item $\left(\frac{d}{2}\right)$ is $1$ if $d$ has residue 1 modulo 8
and $-1$ if it has residue 5 modulo 8.
\end{enumerate}

For its computation one can use the celebrated {\it Law of
Quadratic Reciprocity}: if $p,q$ are coprime positive odd numbers
then
$$
\left(\frac{p}{q}\right)\left(\frac{q}{p}\right)=(-1)^{\frac{p-1}{2}
\frac{q-1}{2}}.
$$

Here is the list of the discriminants $d$ up to 100 with $h(d) = 1$, see
 \cite{Hua} 
$$
5, 8, 13, 17, 20, 29, 37, 41, 52, 53, 61, 65, 68, 73, 85, 89, 97.
$$
It is believed that there are infinitely many fundamental discriminants with
$h(d)=1$, but it is still an open problem. Notice that for positive
definite forms it is known that there are only 9 fundamental
discriminants with $h(d)=1$ as it was conjectured by Gauss, namely
$$
d=-3, -4, -7, -8, -11, -19, -43, -67, -163.
$$

 In general if $h(d) >1$ the right-hand side of the formula
(\ref{Nn}) gives the total number of representations of $n$ by all
non-equivalent forms with discriminant $d.$ An interesting case is when the ideal class number of $d$ is 1 but $h(d)=2$. In that case we have only two non-equivalent forms with discriminant $d$:  $Q$ and $-Q$
and the formula (\ref{Nn}) gives the number of the solutions of the equation $|Q|=n.$
The first corresponding discriminants are:
$$
12, 21, 24, 28, 32, 33, 44, 45, 48, 56, 57, 69, 72, 76, 77, 80, 84, 88, 92, 93
$$
(see \cite{Hua}).
The only discriminants $<100$ not listed in either table above are $40, 60, 85, 96$ with $h(d) = 2,4,2,4$, 
respectively.
Note that most of these discriminants $d$ are not of the form $D = t^2 - 4$, but they can still
be obtained from $A \in SL_2({\mathbb Z})$ because $D/d$ may be an arbitrary  square.
 
Now we are ready to describe the multiplicities of the eigenvalues
$\mathcal E_{l, [\gamma]}.$ First we should take into account that
the gluing map $A \in SL(2,{\mathbb Z})$ and the corresponding form
$Q = Q_{A^*}$ may be non-primitive. Let us define the positive
integers $r=r(A)$ and $l=l(A)$ from the relations $A = A_0^r$ and
$Q_{A^*} = l \hatQ,$ where $A_0 \in SL(2,{\mathbb Z})$ and $\hatQ =
\hatQ_{A^*}$ are primitive.

\begin{theorem}
The multiplicity $m$ of the eigenvalue $\mathcal E_{l, [\gamma]}$
of the Laplace-Beltrami operator $\Delta$
for generic values of the parameters in the metric (\ref{metric}) 
is
$$
m(\gamma) = 2 r(A) N_{Q^*}(n),
$$
where $n = Q^*(\gamma) =
l(A)^{-1} Q_{A^*}(\gamma).$ When the discriminant $d$ of the form $Q^*$
has class number 1 and $n$ coprime $d$ then $N_{Q^*}(n)$ can be computed using the
formula (\ref{Nn}).
\end{theorem}

\noindent
{\bf Examples.} 

1. Let
$$
A=\left(\begin{array}{cc} 2 & 1
\\ 1 & 1
\end{array}\right)
$$
be the so-called {\it cat-map.} Then $A^* =
A$ and $Q_A = Q_{A^*} = -(x^2 -xy -y^2)$ are both primitive. The
discriminant $D=d=5$ has class number 1, so one can use the formula
(\ref{Nn}) to compute the multiplicity of the corresponding
$\mathcal E_{l, [\gamma]}.$ One can check that this leads to the
formula
$$
m = 2(N_{\pm 1}(n) - N_{\pm 2}(n)),
$$
where $N_{\pm 1}(n)$
and $N_{\pm 2}(n)$ are the numbers of divisors of
$n=Q_A(\gamma)$ which have respectively the residues $\pm 1$ and
$\pm 2$ modulo 5.

This example shows that the multiplicities of the eigenvalues can
be as big as we like: for example for $n = 11^M$ the multiplicity
is $M+1$; for $n$ a product of $M$ distinct primes all $\pm 1 \bmod 5$ the
multiplicity is $2^M$.

2. The matrix $$
A=\left(\begin{array}{cc} 
1 & 3\\ 
3 & 10
\end{array}\right)
$$
corresponds to $d = 13$ with $D = 3^2 d$ and $Q_{A^*} = -3(x^2 - 3 xy - y^2)$ so $l(A) = 3$.

3.  For $$
A=\left(\begin{array}{cc} 
5 & 2\\ 
2 & 1
\end{array}\right)
$$
we have $d = 8$ with $D = 2^2 d$ and $ Q_{A^*} = Q_{A} =-2(x^2 - 2 xy - y^2)$ so $l(A) = 2$.
This example shows that $l(A)$ in general is not directly related to the largest square divisor of $D$.

4. For $$
A=\left(\begin{array}{cc} 
1 & 3\\ 
1 & 4
\end{array}\right)
$$
$d = D = 21$ and $Q_{A^*} = -(3x^2 + 3 xy  -  y^2)$, $l(A) = 1$. 
Here $h(d) = 2$, but the non-equivalent forms
simply differ by a sign. 

5. The matrices $$
A_1=\left(\begin{array}{cc} 
1 & 6\\ 
6 & 37
\end{array}\right)
\text{ and }
A_2=\left(\begin{array}{cc} 
7 & 18\\ 
12 & 31
\end{array}\right)
$$
correspond to $d = 40$ with $D = 6^2 d$, $l(A_i)=  6$.
Then $Q_{A_1^*} = -6( x^2 + 6 xy - y^2)$ and
$Q_{A_2^*} = -6 (3 x^2 + 4 xy - 2 y^2)$ are the 
two corresponding (non-trivially) non-equivalent forms.

{\bf Remark.} In the case when $h(d)$ is larger than 1 in general we do not have a simple formula for the multiplicities for a particular $Sol$-manifold but only for the disjoint union of $Sol$-manifolds with non-equivalent forms of given discriminant $d.$ 

The fact that the multiplicities are large and not sensitive to
the change of the parameters in the metric seems to be remarkable.
A possible explanation of the rigidity of multiplicities for
$Sol$-manifolds is in the hyperbolicity hidden in the topology
of the manifolds.

{\bf Remark.} The same numbers $N_Q(n)$ appear in the harmonic analysis
on $Sol$-manifolds as the multiplicities of the irreducible $Sol$-representations
in $C^{\infty} (M^3_A)$ (see Chapter 1 in \cite{Brezin}).
Although this fact has a similar origin it does not explain the degeneracy of the 
spectrum of $\Delta$. In fact one can check that that the same degeneracy holds for a more general class of the metrics on $M^3_A$:
\begin{equation}
\label{metric2}
 \dee s^2=\alpha(z) \dee x^2+ 2\beta(z)\dee x\dee y +\gamma(z)\dee y^2 +
\dee z^2
\end{equation}
where
$$
\left(\begin{array}{cc} \alpha(z) & \beta(z)
\\ \beta(z) & \gamma(z)
\end{array}\right)=
 \exp(-zB)^\top
\left(\begin{array}{cc} \alpha_0(z) & \beta_0(z) \\ \beta_0(z) & \gamma_0(z)
\end{array}\right) \exp (-zB).
$$
Here $\exp B=A$ and $\alpha_0(z), \beta_0(z), \gamma_0(z)$ are
arbitrary real 1-periodic functions with the only condition that
the form $\dee s^2=\alpha_0(z) \dee x^2+ 2\beta_0(z) \dee x \dee y +\gamma_0(z) \dee y^2$ is positive
definite for all $z$. The $Sol$-invariant metrics (\ref{metric}) correspond to the case when these functions are constant.
The degeneracy in this case follows again from the separation of variables and thus is not directly related to the $Sol$-invariance.

It is interesting to compare the $Sol$-case with the spectra of flat
tori $T^2.$ It is easy to see that in the last case the answer will depend
drastically on the metric parameters (or equivalently, on the
geometry of the basic parallelogram). For example, if it is a
square then the spectrum up to a multiple is given by the values
of the standard quadratic form $n = x^2 + y^2,$ and the
multiplicity are given by the  Gauss' famous formula
$$
m = 4(N_1(n)
- N_3(n)),
$$
where $N_1(n)$ and $N_3(n)$ are the numbers of the
divisors of $n$ with the residues $1$ and $3$ modulo 4
respectively. If however the basic parallelogram is generic then
all multiplicities are 2 (which is due to the central symmetry of
the problem).
 
\section{Semiclassical analysis and Weyl's law}

It is instructive to see how our exact calculation of the spectrum agrees with the famous Weyl's law \cite{Weyl}, which says that for a quantum system the number $N(\Lambda)$ of the eigenvalues $\mathcal E \leq \Lambda$  for large $\Lambda$ asymptotically is equal (up to a factor $(2\pi)^{-n}$) to the volume of the domain in the classical phase space with the energy less than $\Lambda$.
For our Laplace-Beltrami operator (\ref{3'}) this means that
\begin{equation} 
\label{WeylfromVol}
N(\Lambda) \sim  \frac{4}{3} \pi \Lambda^{3/2}  \frac{Vol(M_A^3)}{(2\pi)^3} = \frac{4}{3} \pi \Lambda^{3/2}  \frac{{\Cal A}(T^2)}{(2\pi)^3},
\end{equation}
where $Vol(M_A^3)$ is the volume of our $Sol$-manifold (which equals the area of the fibre ${\Cal A}(T^2)$ since the length in $z$-direction was assumed to be 1).

Let us count now the eigenvalues $\mathcal E$ using the results of section 5. 
Let us assume for simplicity that $\cos \theta =0,$ so besides the trivial part they
coincide with the eigenvalues of the Mathieu operator
\begin{equation}
\mathcal M = -\frac{\dee^2}{\dee z^2}  + |\nu| \cosh 2\mu z 
\label{Math}
\end{equation}
where as before 
\begin{equation}
\nu = \frac {8 \pi^2 Q }{ \sqrt{D} {\Cal A}(T^2) \sin\theta}
\label{nu}
\end{equation}
 and $Q$ is the corresponding binary quadratic form.

First of all let us use the well-known fact from number theory (see e.g. \cite{Hua}) that
for large $Q_0$ the number of lattice points (modulo $A$) with values of $|Q|$ less than $Q_0$ 
is proportional to the area of the fundamental domain up to $Q_0$ :

\[
   M(Q_0) \sim 4 \frac{ \mu }{\sqrt{D} } Q_0\,.
\]

The factor of four counts lattice points related by the symmetry given by changing 
the sign of both $p_u$ and $p_v$ and also accounts for the states in the quadrants
where $Q$ is of opposite sign.

For fixed value of $Q$ (hence $\nu$) there is a whole line of eigenvalues of the 
Mathieu operator (\ref{Math}). 
The number of these eigenvalues up to energy $\Lambda$ for large $\Lambda$
is given asymptotically by the {\it action integral} 
\[
  I( \Lambda, Q) = \frac{1}{2\pi} \oint \sqrt{  \Lambda - |\nu| \cosh2 \mu z)} dz \,,
\]
which is of course the area of the domain in the phase plane with energy less than $\Lambda$
divided by $2 \pi$. This can be simplified to 
\[
  2\pi \mu I =  \sqrt{ \Lambda} \oint \sqrt{ 1 - g \cosh 2\zeta } \dee \zeta, \quad g = |\nu|/ \Lambda \,.
\]
With $\xi = \cosh(2 \zeta)$ this becomes a standard elliptic integral (see e.g. \cite{WW})
\begin{equation} 
\label{act}
\frac{2 \pi \mu }{ \sqrt{ \Lambda} }  I =2 \int_1^{1/g} \frac{ \sqrt{ 1 - g \xi} }{\sqrt{\xi^2 - 1}} \dee \xi = 
   4 \sqrt{1 + g} (K(k) - E(k)),\quad
  k^2 = \frac{1-g}{1+g}.
  \end{equation} 
Let us denote this expression $f(g).$

Thus we see that the total number of states up to energy $\Lambda$ is 
\[
  N(\Lambda) \sim  \int  M'(Q) I(\Lambda, Q) \dee Q =  2 \frac{ \Lambda^{3/2}}{C \mu \pi}  \int_0^1 \frac{\mu}{\sqrt{D}} f(g) \dee g,
\]
where $C = \frac {8 \pi^2}{ \sqrt{D} {\Cal A}(T^2) \sin\theta}.$
The integral over $g$ is best performed by treating it as a double integral over $g$ and 
$\xi$. Introducing $\eta = g \xi$ and performing the $\eta$-integral first gives
\[
\int_0^1 f(g) \dee g =  \frac43 \int_1^\infty \frac{\dee \xi}{\xi \sqrt{\xi^2 - 1}}  = \frac23 \pi \,.
\]
Thus we have
\begin{equation} \label{WeylfromI}
   N(\Lambda) \sim \Lambda^{3/2} \frac{ {\Cal A}(T^2)}{(2\pi)^3} \frac{4 \pi}{3} \sin\theta \,,
\end{equation}
which agrees with Weyl's formula (\ref{WeylfromVol}) when $\theta = \pi/2.$

For general $\theta$ we have $\mathcal E = \lambda - \nu \cos\theta$. 
In that case we need to compute
\[
  X_\pm  = 
  \int \int \sqrt{ 1 - g ( \cosh 2 z \pm \cos \theta) } \dee z \dee g 
\]
over the domain $ 0 \le g \le 1/(1 \pm \cos\theta)$ and $|z| \le z_0$ where $z_0$ is the smallest
positive root of the integrant. The transformation 
\[
   \eta = g (\cosh 2 z + \cos \theta), \quad 
   \xi = \cosh 2 z
\]
folds the integration region to the rectangle $\xi > 1$ and $ 0 \le \eta \le 1$ and
the integral becomes 
\[
    X_\pm = 2 \int\int \frac{ \sqrt{1-\eta}}{(\xi \pm \cos\theta )\sqrt{\xi^2 - 1}} \dee \eta \dee \xi \,.
\]
The integral over $\eta$ is easily done as before while the integral over $\xi$ evaluates to 
\[
   X_\pm = \frac{4}{3}  \frac{ \frac{\pi}{2} \pm (\theta - \frac{\pi}{2}) }{\sin\theta} \,.
\]
Hence the sum of the contributions
from the two cases of positive and negative $Q$ is
\[
  \sin \theta ( X_+ + X_-) = \frac{4}{3} \pi 
\]
as before. This computation shows that $\theta$ determines the relative number of states between
the regions with positive and negative $Q$, namely $X_+/X_- = \theta/(\pi-\theta)$.

Let us look what this calculation gives for the first eigenvalues of $\Delta.$
There are two opposite cases depending on whether geometric parameter
$A = {\Cal A}(T^2) \sin\theta$ is small or large.
Let us assume again for simplicity that $\theta = \pi/2,$ then $A$ is simply the area of the fibre.

The small $A$ corresponds to the "rope-like" $Sol$-manifolds.
In this case the first eigenvalues
are "trivial": $\mathcal E_k = 4 k^2 \pi^2, \quad k=0,1,2... $ which correspond to $Q=0.$

The second case when $A$ is large is more interesting. In that case the parameter $\nu$
in the Mathieu operator is small. The action integral (\ref{act}) for small $\nu$ has the asymptotics
\footnote{The fact that the action diverges logarithmically for $Q \to 0$ seems to be surprising.
To our knowledge this is the first example of Liouville integrable system on a compact manifold for
which the action diverges on approach of a singular level (with energy fixed). }
$$
I \sim \frac{\sqrt \Lambda}{\mu \pi} \ln \frac{\Lambda}{|\nu|}.
$$

\begin{figure}
\centerline{\includegraphics[width=6cm]{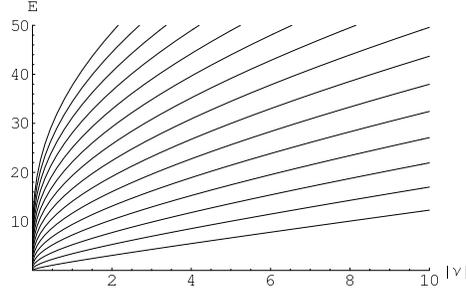}}
\caption{First 15 states of the cosh-Mathieu equation in dependence of the parameter.} \label{speccosh}
\end{figure}

This suggests the following asymptotics for the eigenvalues for small $\nu$:
$$
\Lambda_k = \frac{(\mu \pi k)^2}{(\ln |\nu|)^2}, k=0,1,2... .
$$
Note that although $\Lambda_k \to 0$ as $\nu \to 0$ the decay is slower than any power of $\nu$.
The first 15 states are shown in fig.~\ref{speccosh}.

For the corresponding $Sol$-manifolds this gives the following behaviour of the first eigenvalues
\begin{equation} 
\label{smallnu}
\mathcal E_{k ,[\gamma]} \sim \frac{(\mu \pi k)^2}{(\ln |C  Q_{A^*}([\gamma])|)^2}, \quad k=0,1,2,... ,\quad [\gamma]  \in \Gamma^*\setminus \{0\} / {A^*} 
\end{equation}
for small $C = \frac {8 \pi^2}{ \sqrt{D} {\Cal A}(T^2) \sin\theta}.$
If we order these eigenvalues $\mathcal E_{0}=0 \le \mathcal E_{1}  \le \mathcal E_{2}, \dots $
then we have
\begin{equation} 
\label{first-smallnu}
\mathcal E_{j} \sim  \frac{(\mu \pi)^2}{(\ln C)^2} (1 - 2 \frac {\ln Q_j}{\ln C}), \quad j = 1, 2, \dots,
\end{equation}
where $Q_0 = 0 \le Q_1\le Q_2, \dots $ are positive values of the form $Q_{A^*}$ listed in increasing order and we have assumed that $Q_j \ll \frac{1}{C}$. In particular,
\begin{equation} 
\label{first-smallnu-Q}
\mathcal E_{1} \sim   \frac{(\mu \pi)^2}{(\ln C)^2},\quad \quad \frac{\mathcal E_{j} - \mathcal E_{1}} {\mathcal E_{i} - \mathcal E_{1}} \sim  \frac{\ln (Q_j / Q_1)}{\ln (Q_i/Q_1)},
\end{equation}
so when $C$ is small we "see" the values of the quadratic form $Q_{A^*}$ straight from the spectrum.

Note that the question about the next order term in Weyl's law is non-trivial. For the simpler case of $Nil$-manifolds some results in this direction can be found in \cite{PT}.

One can also look at the corresponding Minakshisundaram-Plejel asymptotic expansion, which is an important characteristics of the spectra (see e.g.\cite{G}). In particular the second coefficient in this expansion is proportional to the integral of the scalar curvature $K$. A straightforward calculation shows that for the $Sol$-manifold $M^3_A$ the principal sectional curvatures are $\pm \sin2\theta \log^2\lambda$ and $-2(\sin\theta \log\lambda)^2$, so 
$$K = -2(\sin\theta \log\lambda)^2$$
and thus is always negative.

\section{Spectral Statistics}

The spectral statistics of integrable and chaotic systems is quite different, see, e.g., \cite{Berry-Tabor, BGS,DG}. As we have seen the geodesic flow on $Sol$-manifolds has
 properties of both, integrable and chaotic systems. Therefore it is a natural question
what the spectral statistics of the $Sol$-manifolds is like.  Note that according to the Berry-Tabor conjecture \cite{Berry-Tabor} integrable systems should have Poisson distributed level spacing.
We are going to show that this is not the case for $Sol$-manifolds.
 
The reason is the high multiplicities of the eigenvalues.
Indeed, for the simplest positive quadratic form $Q_0=x^2 + y^2$
the classical result due to E. Landau says that the number of
integers up to a number $K$ represented by this form grows as
${K}/{\sqrt{\log K}}$.
If there would be no degeneracies then this number 
would grow like the area of the fundamental region, which is proportional to $K$. 
This means that most of the level spacings of the values of $Q_0$ are zero.

\begin{figure}
\centerline {\includegraphics[width=1.8in]{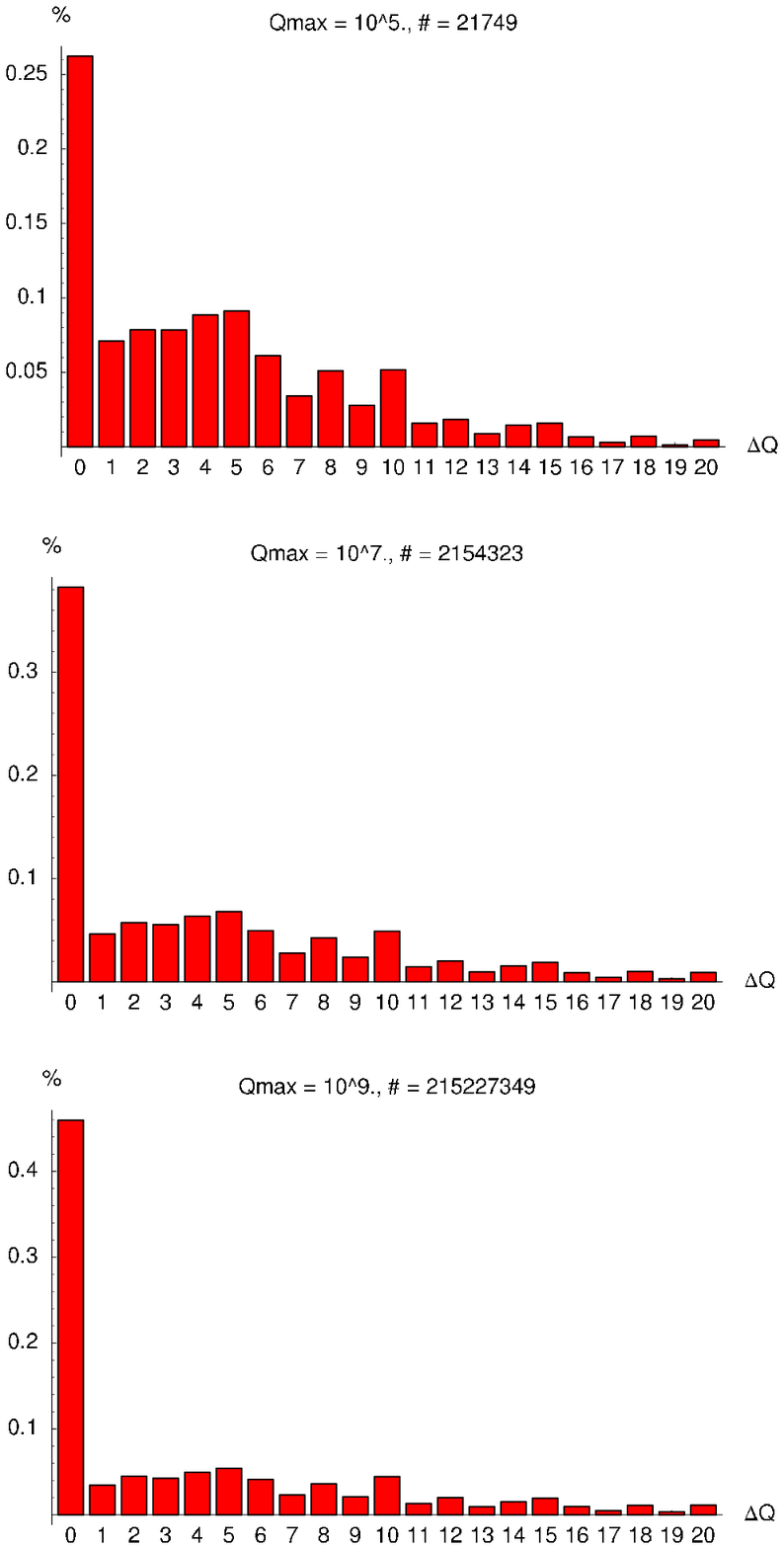} \hspace{1cm} 
\includegraphics[width=1.8in]{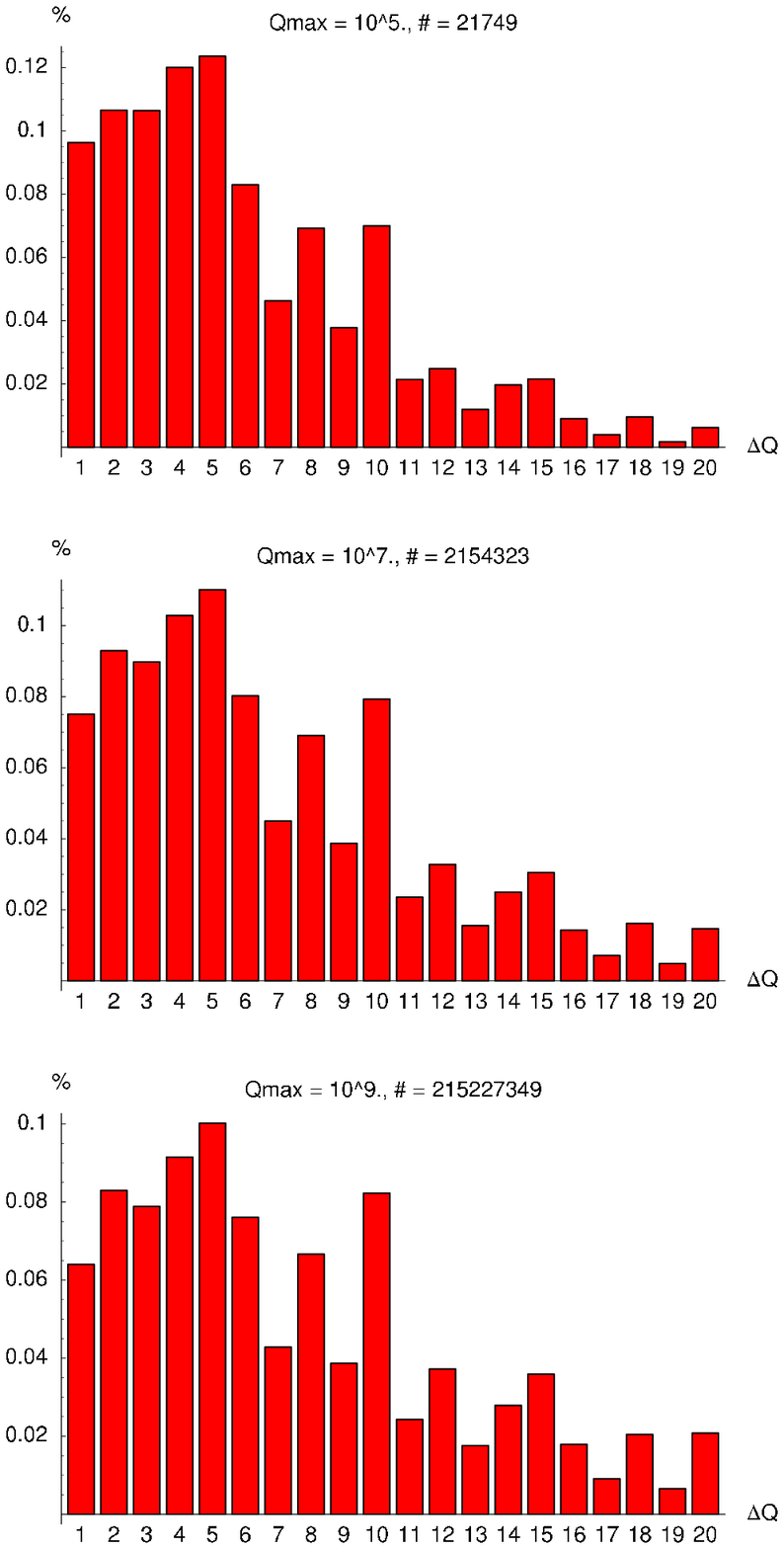}}
\caption{Level spacing statistics of the indefinite binary quadratic form $Q(x,y) = -x^2 + xy + y^2$
(left) with degeneracies removed (right)}
\label{bacha1}
\end{figure}


According to P. Sarnak \cite{Sarnak1, SarnakPC} a similar fact is true for indefinite forms as well, namely 

\vspace{1ex}
\noindent {\em The number of positive integers up to $K$ that can be represented by a given 
indefinite quadratic form $Q$ grows not faster than $O(K / \sqrt{ \log K})$.}
\vspace{1ex}

Combining this with the results of section 5 we have the following

\begin{theorem}
The level spacing distribution for the spectrum of $Sol$-manifolds $M^3_A$ is not Poisson and hence the Berry-Tabor conjecture does not hold in this case. 
\end{theorem}

This is particularly interesting because this statement is  not sensitive 
to change of the metric in the $Sol$-class (\ref{metric}) (or even more general class (\ref{metric2})).
 
Let us illustrate this in the example of the cat-map $A$.
In fig.~\ref{bacha1} (left) the level spacing statistics for the indefinite binary quadratic form
$Q_A= -x^2 + xy + y^2$  is shown for three different values of $Q_\text{max}$.

Since the cat-map is the product of two involutions, there is a simple reflection 
symmetry in the lattice, which causes almost all states to be at least twofold degenerate.
This discrete symmetry needs to be factored out before the level spacing statistics can
be studied. The involutions $R_i$ with $R_i^2 = Id$ are
\[
  A = R_2 R_1, \quad 
  R_1 = \begin{pmatrix} 1 & 0 \\ -1 & -1 \end{pmatrix}, \quad
  R_2 = \begin{pmatrix} 1 & -1 \\ 0 & -1 \end{pmatrix} \,.
\]
The fixed line of $R_1$ is the line $y = -x/2$, and factoring the fundamental region
in fig.~\ref{lattice} by $R_1$ simply cuts the fundamental region in half along this
line. Since the values of $Q$ are integers we chose to present the raw level spacing 
statistics, i.e.~without unfolding the spectrum first.
The number of lattice points found up to the corresponding $Q_\text{max}$ in the reduced
fundamental region is given in the heading of each figure, and the ratio approaches 
$\ln \lambda / (2 \sqrt{d})$.
The figures clearly show that the proportion of degenerate levels grows
in agreement with what we said above.
When the degenerate levels are discarded in the statistics fig.~\ref{bacha1} (right) shows
that the distribution appears to converge to some non-universal shape.  We would like to mention here the paper  \cite{Hooley} where the moments of the intervals between the sums of two squares were studied.

\section {Quantum monodromy}

\begin{figure}
\centerline{ \includegraphics[width=6cm]{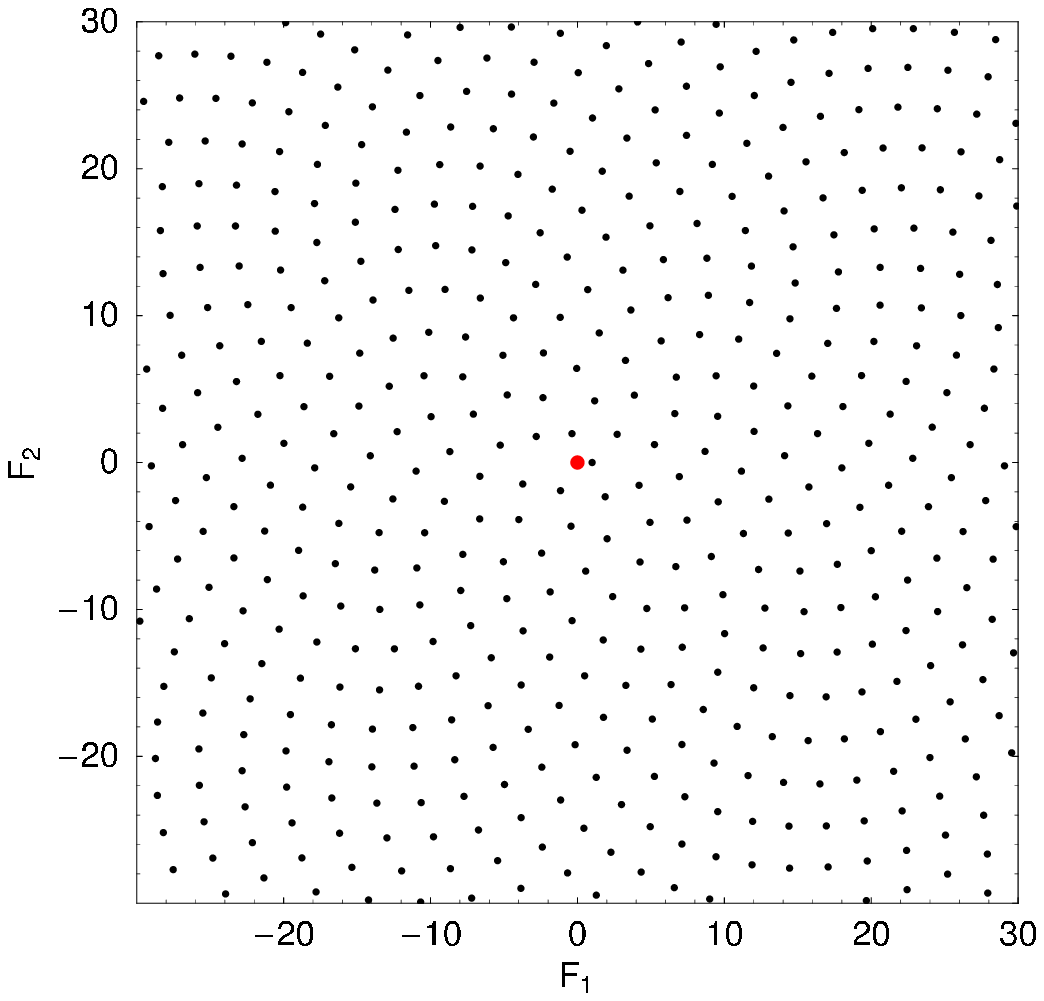} \hspace{10pt} \includegraphics[width=6cm]{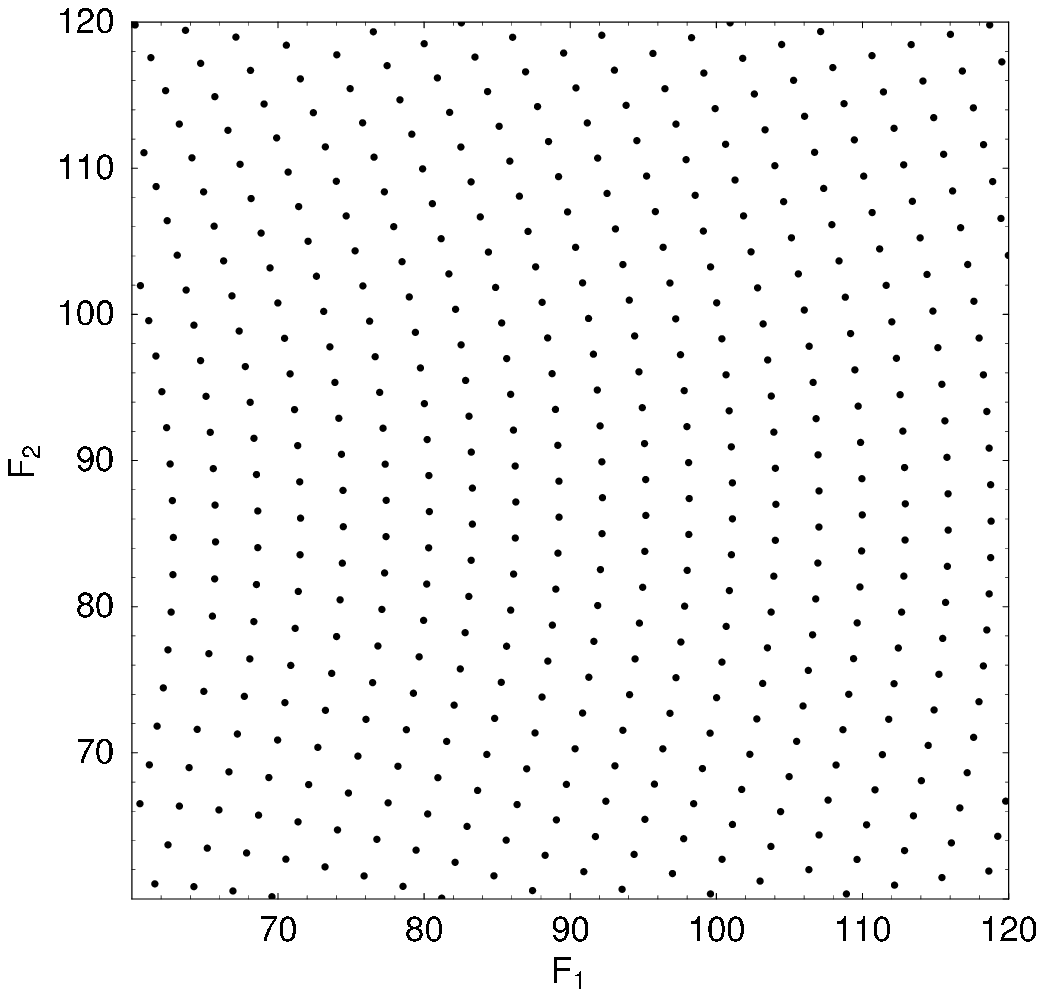} }
\caption{Image of the lattice ${\mathbb Z}^2$ in $(p_x, p_y)$ under the momentum 
map $(F_1, F_2)$ for fixed energy where $A$ is the cat-map. 
Left: origin at the centre. Right: distorted standard lattice away from the origin.} \label{boxflower}
\end{figure}

In view of the previous results the appearance of quantum monodromy in our problem is
quite natural. However there is a problem with this notion in our case which we want to discuss first.

As it was shown in \cite{BT} the geodesic flow on $Sol$-manifolds can not have three analytic integrals (see Section 3 above). A similar fact holds in the quantum case. Namely, one can show that the algebra of the differential operators on the $Sol$-manifold $M_A^3$ commuting with the Laplace-Beltrami operator $\Delta$ (\ref{3'}) is generated by $\Delta$ and $\frac{\partial}{\partial u} \frac{\partial}{\partial v}$.
This means that our quantum problem does not have enough quantum integrals, at least in the class of the differential operators and therefore it is not clear if we can 
apply the rigorous treatment of quantum monodromy from \cite{San}.  

So in this section we will treat the quantum monodromy on $Sol$-manifolds on the intuitive level paying more attention to geometry rather than to analysis.  As we have already shown, the set of eigenfunctions is in a natural one-to-one correspondence with  $\Gamma^*/A^* \times \mathbb N$,  $[\gamma]\in
\Gamma^*/A^*$, $k\in \mathbb N$. 
The fundamental domain of $A^*$ is shown in fig.~\ref{lattice}.
It is natural to represent the
orbit space as a lattice on the cone obtained by gluing the edges
of the fundamental domain of $A^*$ on the plane (more precisely we
should consider four different cones corresponding exactly to four
families of Liouville tori).

Quantum monodromy arises when we pass around the vertex
of the cone. It is clear that the basis of the lattice
will undergo the transformation $A$. On the other hand, nothing
happens to the third direction corresponding to the parameter $k$.
Therefore the quantum monodromy for the $Sol$-manifold $M_A^3$ is given 
by the matrix
$$
\left(\begin{array}{cc} A^* & 0 \\ 0 & 1 \end{array}\right).
$$

We want to emphasize that in this case quantum monodromy has a
purely topological nature. It is determined by the topology of the
underlying manifold, and not by properties of the metric. It does not
depend on the parameters $E,G,F$, moreover the monodromy remains
the same for all metrics of the form
$$
\dee s^2=\dee s_z^2+  \dee z^2
$$
where $\dee s_z^2$ is a flat metric on fibres $T^2_z$ with
coefficients depending on $z$. In the previously known examples (like the geodesic flow on the
3-dimensional ellipsoid of revolution, \cite{WD}) the
metric $g$ played the principal role.

\begin{figure}
\centerline{ \includegraphics[width=6cm]{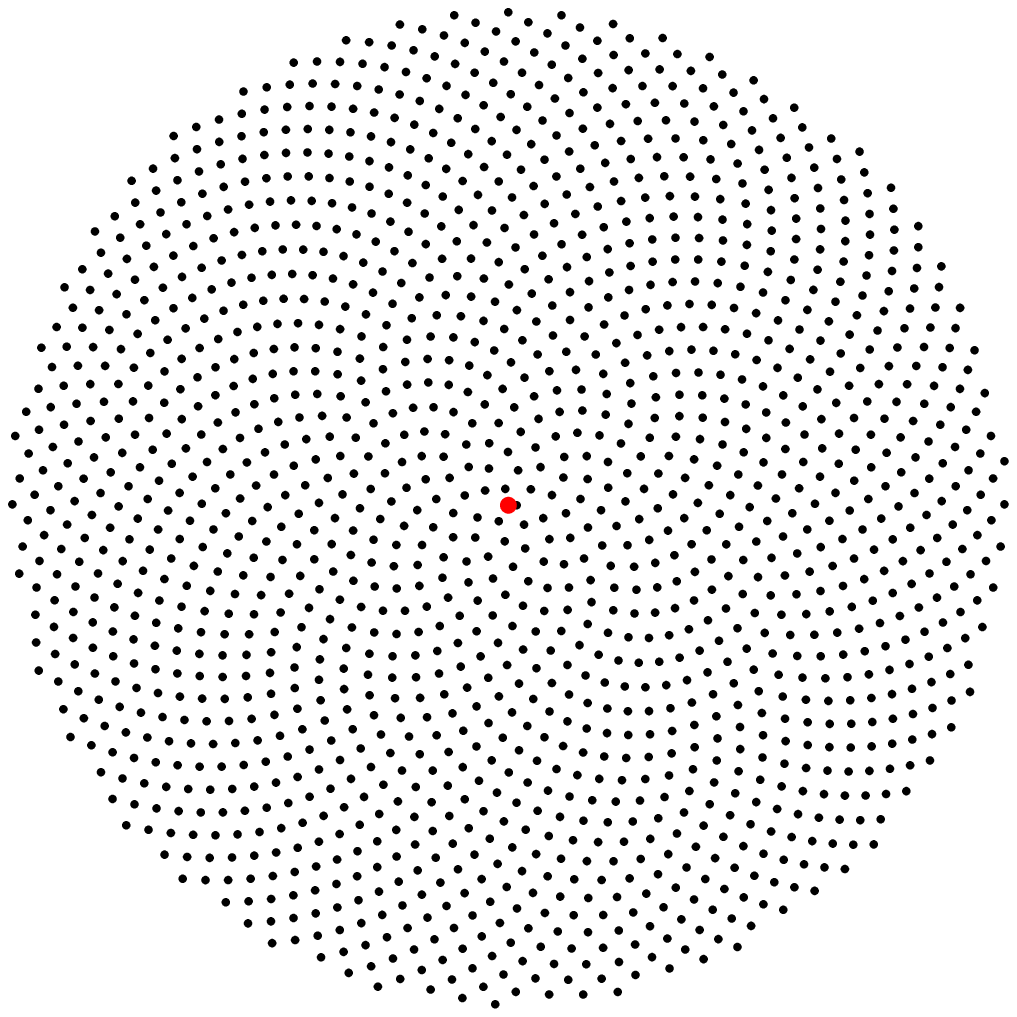} \hspace{10pt} \includegraphics[width=6cm]{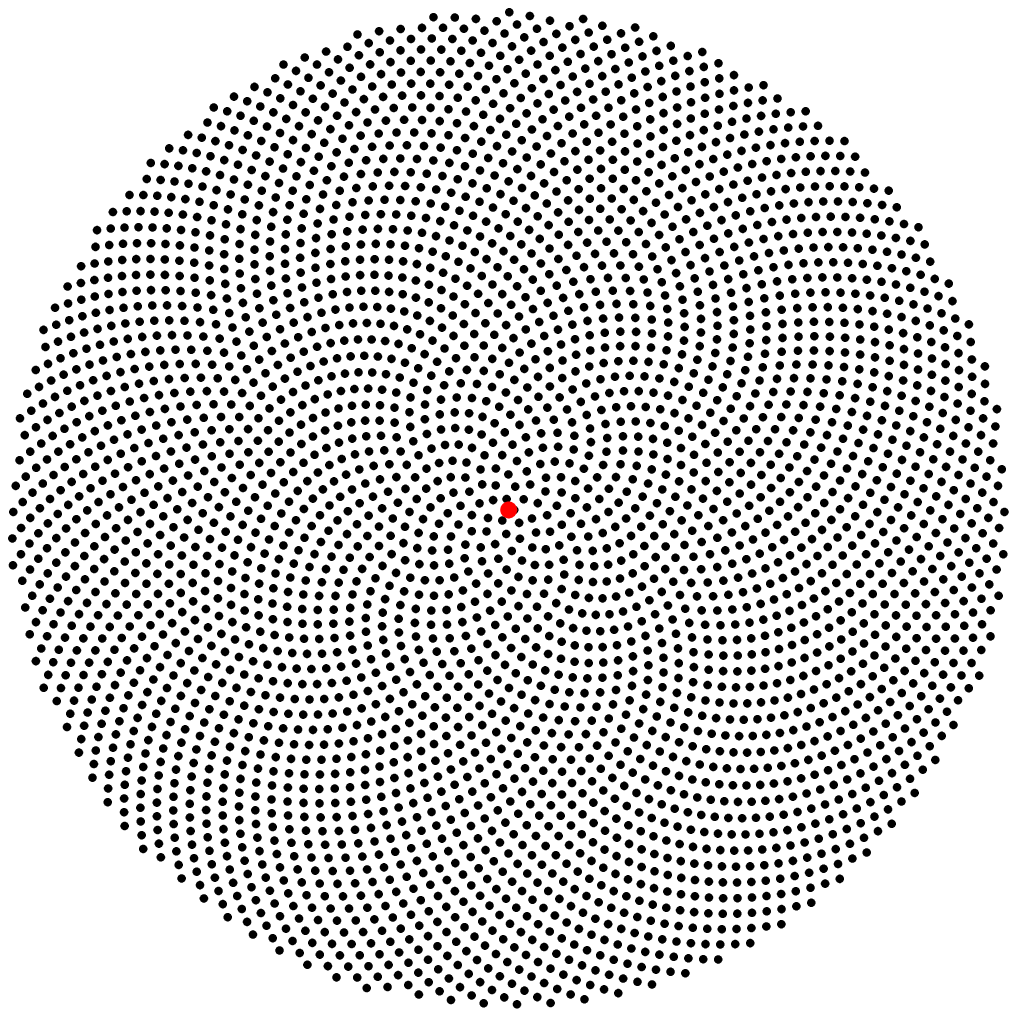} }
\caption{Image of the lattice ${\mathbb Z}^2$ in $(p_x, p_y)$ under the momentum 
map $(F_1, F_2)$ for fixed energy and $A$ the cat-map. Left: $Sol$-flower with $|Q| \le 60^2$. Right: $Sol$-flower with $|Q| \le 90^2$} \label{solflower}
\end{figure}

\begin{figure}
\centerline{ \includegraphics[width=6cm]{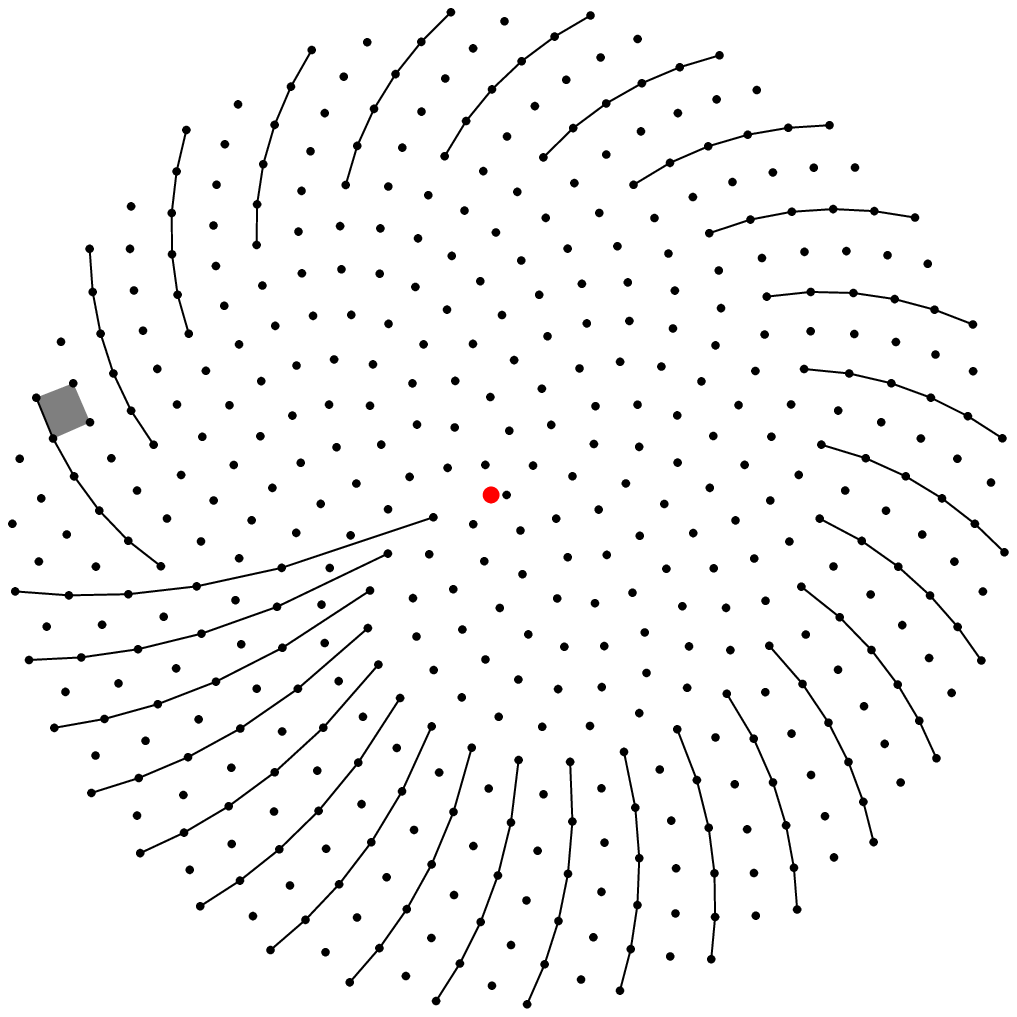} \hspace{10pt} \includegraphics[width=6cm]{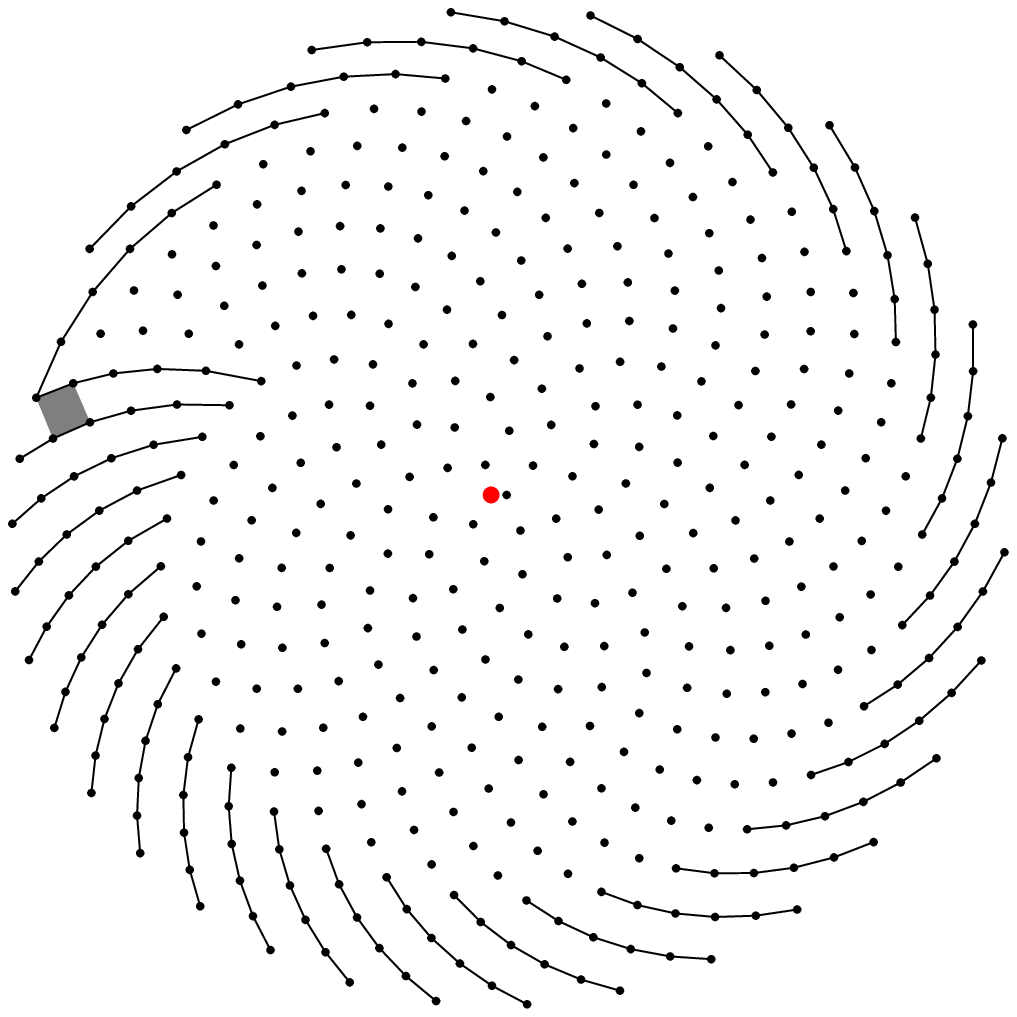} }
\caption{Parallel transport of basic directions in the image of the momentum map.} \label{paralleltransport}
\end{figure}

In figures \ref{boxflower}, \ref{solflower}, and \ref{paralleltransport} 
we demonstrate the quantum monodromy of the $Sol$-manifold
$M_A^3$ related to the cat-map 
$$
A=\left(\begin{array}{cc} 2 & 1
\\ 1 & 1
\end{array}\right).
$$
To make the image of the lattice uniform we have slightly modified the classical integrals 
$f_1, f_2$ from section 3 as follows:
$$
\begin{array}{l}
\label{F}
 F_1= \sqrt{|Q|} \cos 2\pi\beta,\\
 F_2= \sqrt{|Q|} \sin 2\pi\beta,
\end{array}
$$
where $Q = p_u p_v$ and $\beta = \frac{\ln \left|
p_u\right|}{\ln\lambda}$. 
The image of the lattice under the map $F$ we call {\it $Sol$-flower} (see fig.~\ref{solflower}).
Note that an alternative choice $\beta = \frac{\ln \left|
p_v\right|}{\ln\lambda}$ would give a similar picture and the freedom of the rescaling of the eigenvectors $e_u, e_v$ leads simply to a rotation of the plane $(F_1, F_2)$.
Fig.~\ref{boxflower} illustrates that away from the origin the lattice is simply a deformed 
standard lattice.

A nice property of the map $ (p_u, p_v) \rightarrow (F_1, F_2)$
is that it changes the area simply by a constant multiple: it is easy to check that
$$ \dee F_1 \wedge \dee F_2 = \frac{\pi}{\ln\lambda} \dee p_u \wedge \dee p_v.$$ 
It is interesting to mention that the multiplicity problem becomes the standard
"circle problem" if one replaces the square lattice by the $Sol$-flower
(but of course it does not help to compute them).

When a fundamental cell is chosen in the $Sol$-flower as indicated in grey in
fig.~\ref{paralleltransport} the monodromy can be observed as follows:
A line extending a basis vector is parallel transported in the lattice.
After completing a cycle about the origin this direction is changed.
The left picture shows images of the lines $(p_x, p_y) = (30 - 2 l, -j -l)$, $j = 0..27$, $l = 0..5$.
The right picture shows images of the lines $(p_x, p_y) = (30 - l, -j -l)$, $j = 0..30$, $l = 0..5$.
Denote the direction of the line shown in the left part 
of fig.~\ref{paralleltransport} by $e_1$, and the one on the right part by $e_2$.
The preimages of these basis vectors in fig.~\ref{lattice} are $-(2 e_x + e_y)$ and $-(e_x + e_y)$.
Parallel transporting $e_1$ clockwise by increasing $j$ gives $e_1 + e_2$ 
(determining the second row of $A$),
while parallel transporting $e_2$ counterclockwise by decreasing $j$ gives $-e_1 + e_2$
(determining the first row of $A^{-1}$). 
Since $A \in SL(2, {\mathbb Z})$ this determines the cat-map.

\section {Concluding remarks} 

The $Sol$-geometry from a dynamical point of view has the special property  of being on the border between integrability and chaos. 
Integrability is reflected in the solvability of the corresponding group while
the chaos is related to a hidden (partial) hyperbolicity.
This makes the $Sol$-case of particular interest and explains why the geodesic problem on the $Sol$-manifolds has both integrable and chaotic features. As we have seen the quantum case gives a 
new interesting twist to the story by bringing arithmetic into play. 

Atiyah, Donelly and Singer \cite{ADS} considered a more general case of $Sol$-manifolds which are $T^{n+1}$ torus fibres over $T^n.$ Much of our analysis can be generalised to this case as well. The quantum Toda lattice Hamiltonian will appear then as a generalisation of the modified Mathieu
operator. Some very interesting results in the corresponding classical problem were found recently by Leo Butler in \cite{But}.

\begin{figure}
\centerline {\includegraphics[width=3in]{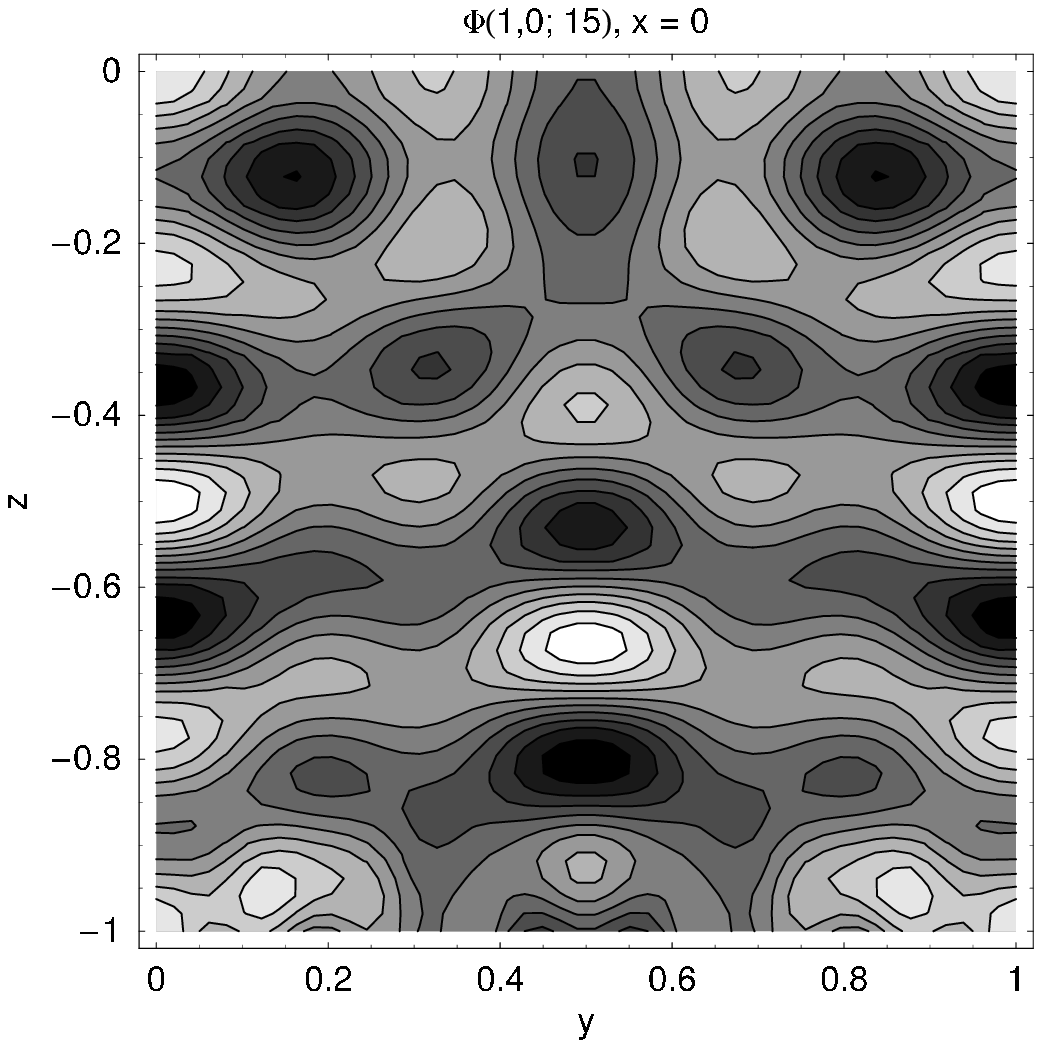} \includegraphics[width=3in]{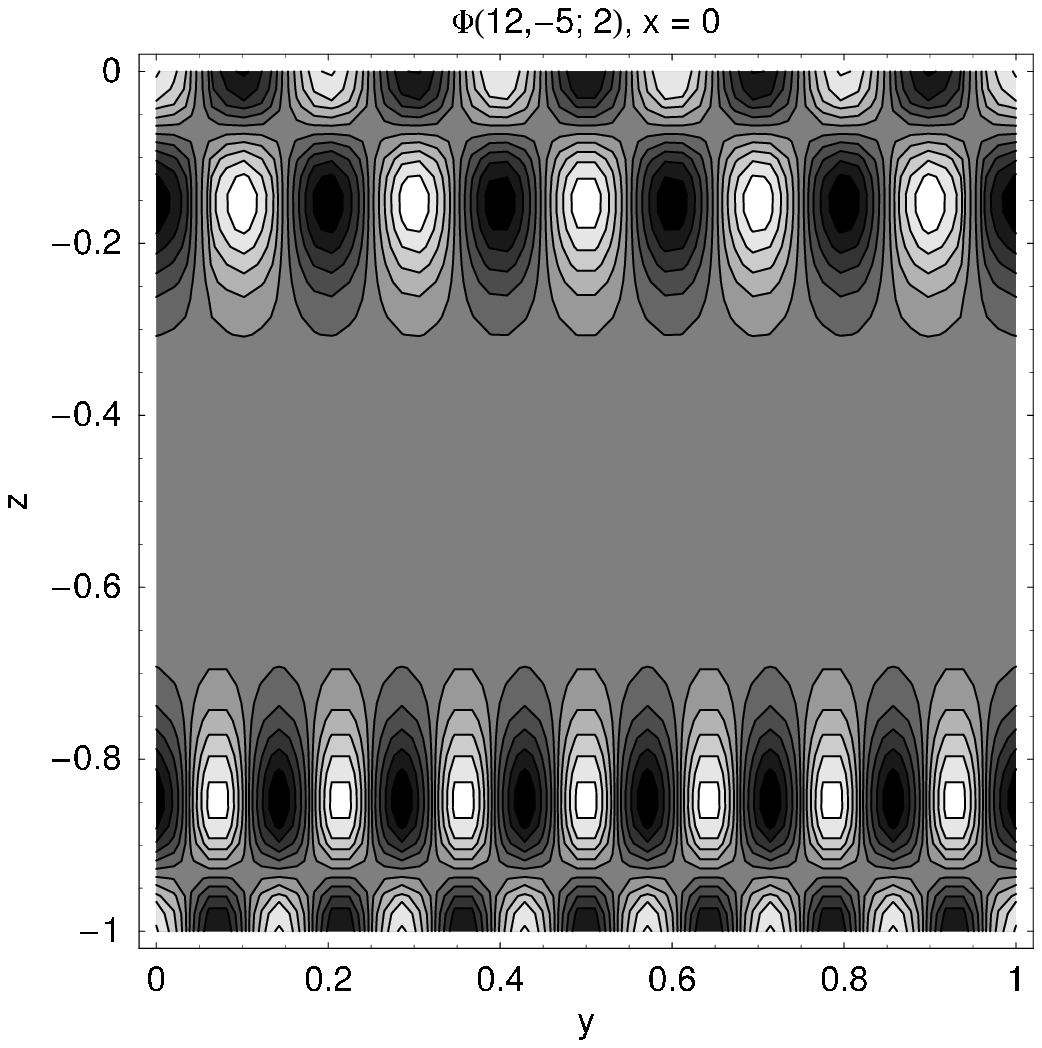} }
\caption{Slice of $\Re(\Phi_{[\gamma], k})$ at $x = 0$ for $\gamma = (1,0), k = 15$ (left)
and $\gamma = (12,-5), k = 2$ (right) for the cat-map. The appearance of the eigenfunction
whose $Q(\gamma)$ is small and thus close to the classical chaos is more irregular.}
\label{chaos}
\end{figure}

It would be also interesting to study in more detail how the chaos (at the degenerate level $Q=0$) of the classical 
system manifests itself in the quantum version. We showed that the spectral statistics 
provides a counterexample to the Berry-Tabor conjecture,
but it cannot be taken as an indicator of chaos.
One simple observation is that the trivial eigenfunctions $\Phi_{0,s}$ are
asymptotically `uniformly distributed' on the manifold. Hence the subset of eigenfunctions
that are associated with the classical chaos are quantum unique ergodic, 
cf.~\cite{JNT}. Already at relatively small quantum numbers it can 
be seen that the nodal lines are more complicated when $Q$ is small, 
see fig.~\ref{chaos}.

\section*{Acknowledgements}

We are very grateful to M. Berry, E. Bombieri, V. Kuznetsov, J. Marklof, A. Pushnitski, P. Sarnak, R. Schubert and  P. Shiu  for very useful and stimulating discussions.

This work has been started in December 2002 when one of us (A.B.) visited Loughborough University.
We are grateful to the London Mathematical Society for the support of this visit.

\end{document}